\documentclass[10pt]{iopart}

\usepackage{graphicx}
\usepackage[utf8]{inputenc}
\usepackage{textcomp}

\usepackage{booktabs}
\usepackage{rotating}
\usepackage{multirow}
\usepackage{subfigure}

\usepackage[sort&compress,comma,square,numbers]{natbib}                                
\usepackage{hypernat}  

\usepackage{url,hyperref}

\usepackage{color}
\usepackage[usenames,dvipsnames,svgnames,table]{xcolor}

\begin{document}

\frenchspacing

\title{Long-term stability of Cu surface nanotips}

\author{V Jansson$^1$, E Baibuz$^1$ and F Djurabekova$^1$}

\address{$^1$ Helsinki Institute of Physics and Department of Physics, P.O. Box 43 (Pehr
Kalms gata 2), FI-00014 University of Helsinki, Finland}
\ead{ville.b.c.jansson@gmail.com}

\begin{abstract}

Sharp nanoscale tips on metal surfaces of electrodes enhance locally applied electric fields. Strongly enhanced electric fields trigger electron field emission and atom evaporation from the apexes of the nanotips. Combined together, these processes may explain electric discharges in form of small local arcs observed near metal surfaces in the presence of electric fields even in ultra high vacuum conditions. In the present work we investigate the stability of nanoscale tips by means of computer simulations of surface diffusion processes on copper, the main material of high voltage electronics.

We study the stability and life-time of thin copper (Cu) surface nanotips at different temperatures in terms of diffusion processes. For this purpose, we have developed a surface Kinetic Monte Carlo model where the jump processes are described by tabulated precalculated energy barriers. We show that tall surface features with high aspect ratios can be fairly stable at room temperature. However, the stability was found to depend strongly on the temperature: 13\;nm nanotips with the major axes in the $\langle 110 \rangle$ crystallographic directions were found to flatten down to half of the original height in less than 100 ns at temperatures close to the melting point, whereas no significant change in the height of these nanotips was observed after 10\;µs at room temperature. Moreover, the nanotips built up along the $\langle 110 \rangle$ crystallographic directions were found significantly more stable than those oriented in the $\langle 100 \rangle$ or $\langle 111 \rangle$ crystallographic directions. 

The proposed Kinetic Monte Carlo model has been found well suited for simulating atomic surfaces processes and was validated against Molecular Dynamic simulation results via the comparison of the flattening times obtained by both methods. We also note that the Kinetic Monte Carlo simulations were two orders of magnitude computationally faster than the corresponding Molecular Dynamics calculations.

\end{abstract}


\submitto{\NT}
\maketitle
\ioptwocol	

\section{Introduction}\label{sec:introduction}

Vacuum arcs are local plasma discharges occurring in high electric fields in a vacuum environment \cite{boxman1995handbook}. The phenomenon has been observed to occur even in ultrahigh vacuum and constitutes a significant problem for high voltage or high gradient electromagnetic field devices. One of the examples of such demanding instruments is linear accelerators with high accelerating gradients, such as Compact LInear Collider (CLIC) under development at CERN \cite{timko2011phd,aicheler2012clic}. The cause of vacuum arcs (or \textit{breakdowns}) are still not properly understood, despite heavy efforts of both experimental and theoretical studies of the phenomenon \cite{boxman1995handbook,anders2012evolution}.

Experimental measurements of the field emission currents from a seemingly smooth surface suggest the existence of short-lived sharp and tall surfaces features (nanotips) that enhance the applied electric field explaining the $\beta$ factor in the Fowler-Nordheim equation for field emission currents \cite{navitski2013field,muranaka2011situ,nagaoka2001field}. These are noticed to be precursors to electric arcs: breakdown events usually occur after electron currents reached the runaway values \cite{muranaka2011situ}. In the study by R\;P\;Little and W\;T\;Whitney, $\sim$2 µm high pre-breakdown nanotips were observed with an aspect ratio of $\sim$10 on copper cathodes, as well as on other metals \cite{little1963electron}. Note that in these experiment the applied electric field was 10 MV/m and the vacuum $10^{-5}$\;Pa (compared to the modern experimental setups, where the values of the electric fields are $\ge 100$ MV/m and the vacuums $\sim$$10^{-7}$\;Pa) \cite{descoeudres2009dc,timko2011energy}. P\;I\;Wang \textit{et al}. have observed the formation of Cu nanorods up to 1 µm tall, with aspect ratios up to $\sim$20, by irradiating a thin Cu film on a Si substrate with an electron beam \cite{wang2004novel}. The authors explain the nanorod-formation by diffusion of Cu atoms on the Si substrate. However, no studies to our knowledge have reported observations of formation of similar nanotips in connection to electrical breakdowns (see, e.g. \cite{shipman2014experimental}) and mostly the field-emitting tips are associated with particles of a foreign origin or metal dust \cite{lagotzky2014enhanced,latham1995high}. Unfortunately this hypothesis does not satisfactorily explain the experimental results obtained recently at CERN \cite{degiovanni2015comparison}, which confirms that the exact nature of the field-emitting nanotips still remains unknown. It is therefore of interest to use multiscale modelling to study how the surface behaves under high electric fields and, if any asperity emerged \cite{pohjonen2013dislocation}, what is the life time of such an asperity after the electric field was removed.

We develop our model for copper (Cu) surface to match the material choice for the accelerating components of CLIC \cite{aicheler2012clic}. At the same time, Cu is a widely used material in many different high voltage devices. To date, there exist many Kinetic Monte Carlo (KMC) models for studying various processes on Cu surfaces, such as film growth due to deposition and surface roughening \cite{hakkinen1993roughening,wang2001kinetic,lam2002competing,zhang2004kinetic,kara2009off,nandipati2012off,tan2005dependence,tan2005pulsed}. Atoms deposited on smooth metal surfaces tend to group together and form islands, which has been clearly observed by scanning tunnelling microscopy \cite{hannon1997surface}. Adatom islands were previously the objects of KMC studies, where the adatom migration energy barriers were either approximated by formulae based on bond-counting arguments \cite{mottet1998monte,larsson2001kinetic}, or calculated on-the-fly by using self-learning algorithms \cite{kara2009off,henkelman2001long,karim2006diffusion}. The former method is fairly inaccurate and may only be effectively applied on relatively smooth surfaces, whereas the latter method is limited by its CPU-intensity. A third method is to estimate the migration energy barriers in an unrelaxed rigid atom lattice using an interatomic potential. This method is fairly approximate since the barriers are dependent on how relaxed the atoms are in the system. The method has, however, been used to study adatom islands \cite{tan2005dependence} and thin film growth by deposition \cite{tan2005pulsed,zhang2004kinetic}.  A bond-counting rule has been applied in a KMC study of surface asperities \cite{frantz2004evolution}, however, it was not validated against experiments nor any other methods.

Surface asperities with high aspect ratios, such as nanotips, have a geometry and behaviour similar to that of nanowires. The properties of the latter have also been studied intensively in both experiments and computer simulations, such as Molecular Dynamics (MD) and KMC. For instance, the MD simulations of W\;Liang \textit{et al}. and H\;S\;Park \textit{et al}. showed that thin nanowires oriented along the $\langle100\rangle$ crystallographic directions may spontaneously reconstruct the shape to align the major axis of the wire along the $\langle110\rangle$ direction \cite{liang2005shape,park2005shape}. Cu nanowires with diameters $\sim$40 nm have also been seen experimentally to break the structural integrity during annealing at temperatures between 670 and 870 K, forming chains of spherical droplets \cite{toimil2004fragmentation}. This effect, known as the ``pearling instability'', was earlier predicted by KMC simulations of Ge nanowires by T\;M\"uller \textit{et al}.\;\cite{toimil2004fragmentation} and was explained by a variant of the Rayleigh instability mechanism. \cite{rayleigh1879capillary} The pearling effect has also been observed in ion-beam irradiated Au and Pt nanowires \cite{zhao2006patterning}. Large nanotips with a radius of about 120\;nm and a height of more than 300 nm were, on the other hand, observed experimentally to be stable at room temperature for more than a month \cite{atamny1995direct}.

In this paper, we study the stability and life-time of thin Cu surface nanotips in terms of diffusion processes of atoms on metal surfaces. For this purpose, we have developed an atomistic Kinetic Monte Carlo model. The model consists of the KMC code \textit{Kimocs}, which is described in section \ref{sec:Kimocs}, and the parameterization of the Cu surface diffusion processes, described in section \ref{sec:parameterization}. The validation of the model is described in  section \ref{sec:validation}, where the processes of flattening of small asperities on $\{110\}$, $\{111\}$, and $\{100\}$ surfaces, and at different temperatures, are compared with corresponding MD simulations. In section \ref{sec:large_tips}, the model is used to investigate the stability and the life time of 13--31\;nm high nanotips on different Cu surfaces and at different temperatures. We also investigate the stability of infinite Cu nanowires to validate the model against the existing experiments. Finally, we discuss the results in section \ref{sec:discussion} and summarize our conclusions in  section \ref{sec:conclusions}.

\section{Computation methods}\label{sec:methods}

\subsection{Kinetic Monte Carlo}\label{sec:Kimocs}

\begin{figure}
 \centering
 \includegraphics[width=\columnwidth]{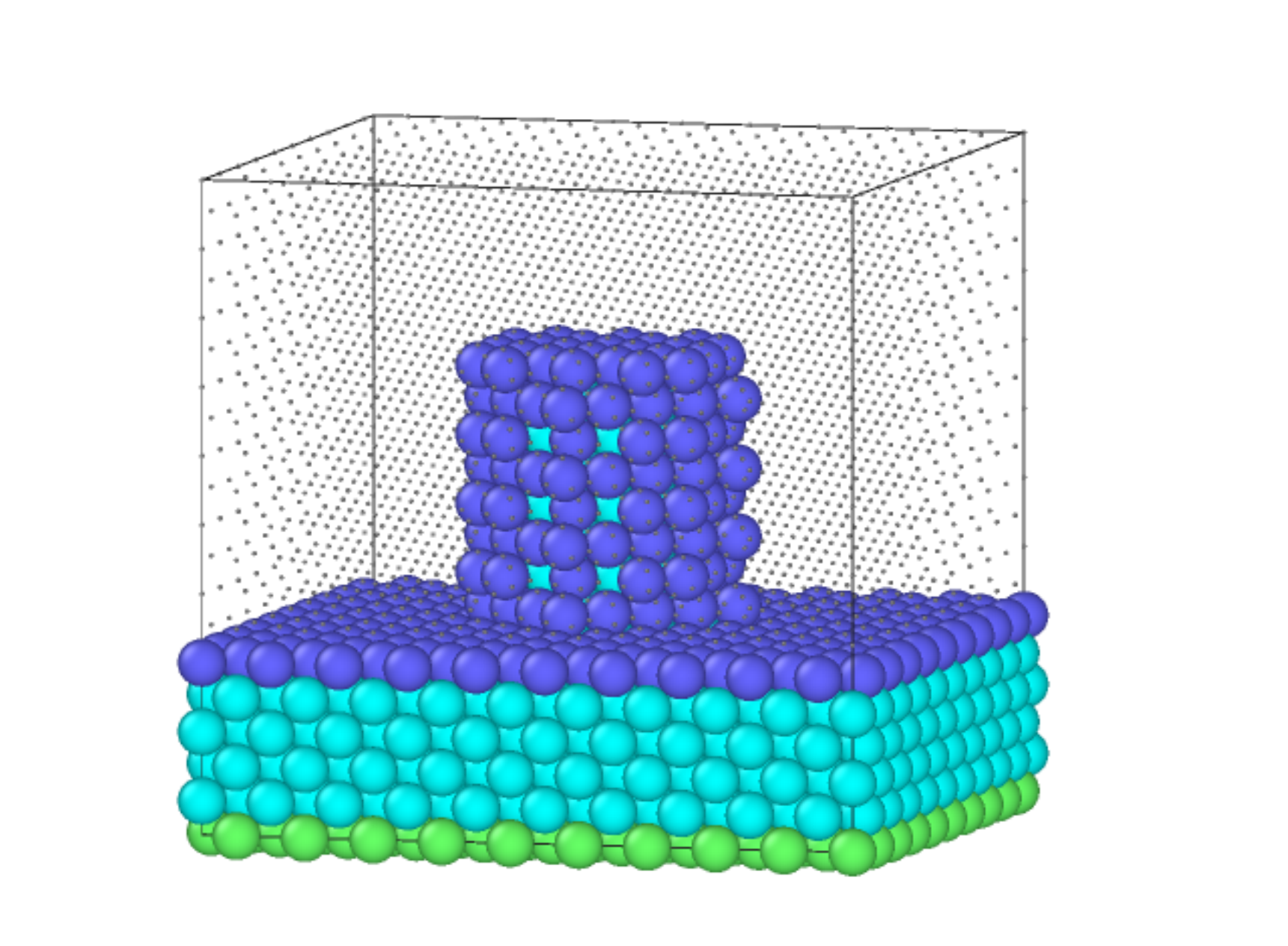}
 \caption{(Colour online) In \textit{Kimocs}, the atoms have to assume the positions on a rigid lattice (dots). Only the atoms with one or more unoccupied first nearest neighbour sites may migrate (blue), as opposed to the atoms that are fully surrounded by other atoms (cyan). Atoms at the lower boundary (green) may be permanently fixed to account for infinite bulk.}
 \label{kimocs_lattice.pdf}
\end{figure}

\begin{figure}
 \centering
 \includegraphics[width=0.6\columnwidth]{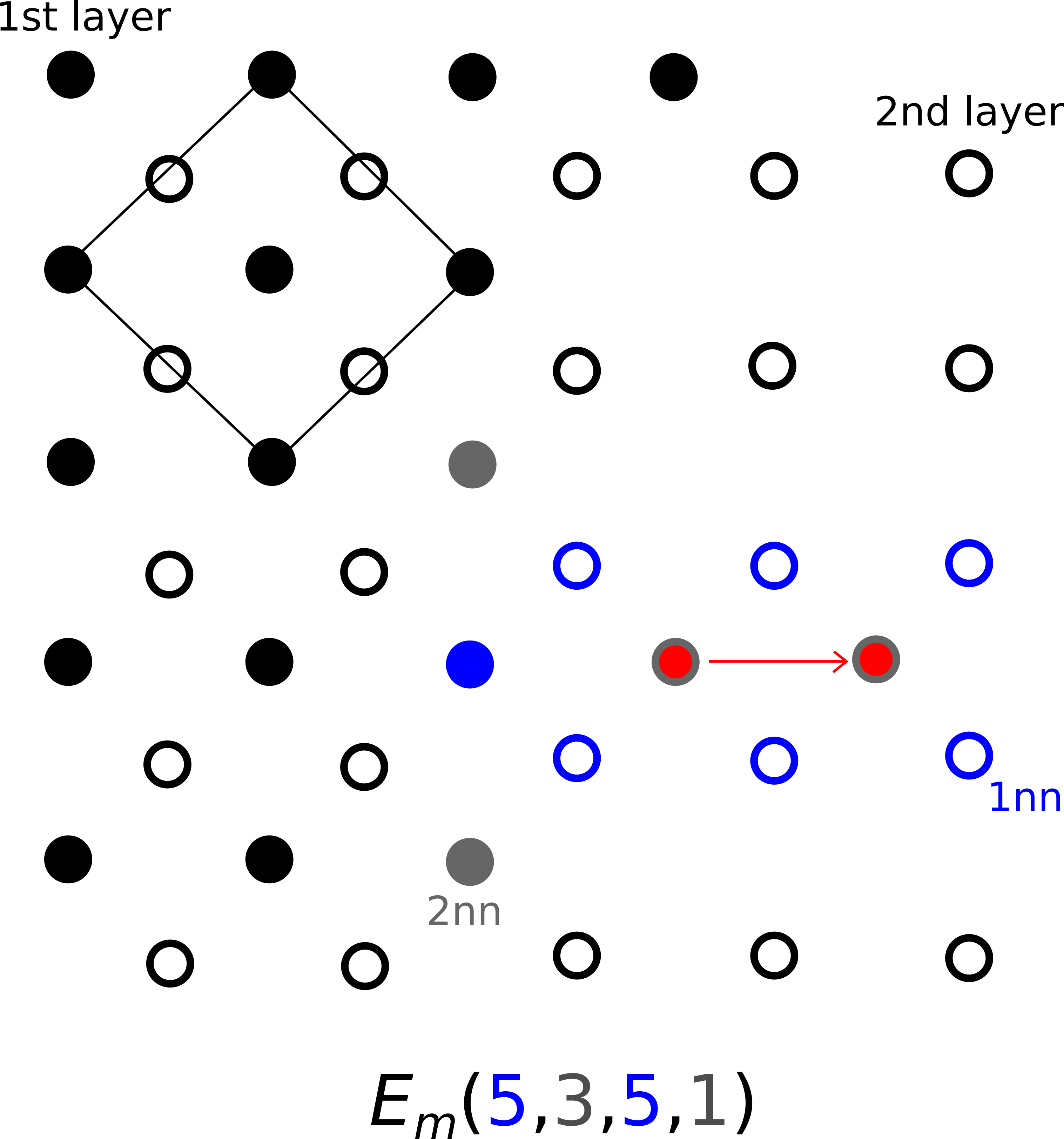}
 \caption{(Colour online) Illustration of the characterization of an atom migration jump on a $\{100\}$ surface in \textit{Kimocs} (between the two sites marked in red). Two atom layers are shown: the first layer (top, shown with filled circles) and the second layer (below, empty circles). The first nearest neighbour (1nn) sites are marked blue and the second nearest neighbour (2nn) sites grey. Note that the migrating atom is counted as one of the 1nn atoms of the final (initially empty) site. To guide the eye, the face-centred cubic (fcc) unit cell is shown by black lines.}
 \label{kimocs_jumps.pdf}
\end{figure}

We present a newly developed atomistic KMC code, \textit{Kimocs}, for simulating the evolution of metal surfaces on the atomic scale. In \textit{Kimocs}, a rigid face-centred-cubic (fcc) lattice is assumed. The lattice spans the whole simulation system, including bulk, surface and vacuum space above the surface (figure \ref{kimocs_lattice.pdf}). At every KMC step, one atom may jump to an unoccupied first nearest neighbour (1nn) lattice site with a precalculated transition rate, in accordance with the general KMC algorithm   \cite{fichthorn1991theoretical,bortz1975new,young1966monte}. Atom jump processes are thermally activated and their transition rates $\Gamma$ are thus given by the Arrhenius formula:
\begin{equation}\label{eq:arrhenius}
\Gamma = \nu\exp \left(\frac{-E_m}{k_B T}\right),
\end{equation}
where
\begin{itemize}
 \item $\nu$ is the attempt frequency of the process
 \item $k_B$ is the Boltzmann constant
 \item $T$ is the temperature of the system
 \item $E_m$ is the migration energy (or activation energy) of the atom to move from one lattice point to another
 \end{itemize}
 In \textit{Kimocs}, all atom jump processes are characterized by the number of the first nearest neighbour (1nn) and the second nearest neighbour (2nn) atoms in the three-dimensional space of the initial and final sites, as shown in figure \ref{kimocs_jumps.pdf}. We denote the number of 1nn and 2nn atoms of the initial site as $a$ and $b$, respectively, and the corresponding numbers for the final site as $c$ and $d$. Then, the migration energy of the jump process is described by four indices: $E_m(a,b,c,d)$. In the fcc lattice, the indices $a$ and $c$ are between 0 and 12, whereas the indices $b$ and $d$ are between 0 and 6. The values of $E_m(a,b,c,d)$ are precalculated and tabulated (section \ref{sec:parameterization}).

In \textit{Kimocs}, the fcc lattice, and thus the atoms, may be arranged in three different orientations, allowing for three different surfaces: $\{100\}$, $\{110\}$, and $\{111\}$. The boundary conditions may be periodic in all three directions: $x$ and $y$ (lateral directions), and $z$ (upwards). With non-periodic boundaries in the $z$ direction, the atoms that reach the upper boundary are removed, whereas the atoms in the bottom layer are assumed to belong in the bulk and thus fixed.

The time increment at each KMC step is calculated according to the residence time algorithm, \cite{bortz1975new}
\begin{equation}\label{eq:residence_time}
 \Delta t = \frac{-\log u}{\sum_i \Gamma_i},
\end{equation}
where the sum is taken over all possible events $i$ at every simulation step and $u \in (0,1]$ is a uniform random number.

\subsection{Nudged Elastic Band}

For calculating the atom migration energy barriers, the minimum energy path was found by using the Nudged Elastic Band (NEB) method \cite{mills1994quantum,mills1995reversible} with the MD code PARCAS \cite{nordlund1998defect,ghaly1999molecular,nordlund1995molecular}. The atomic systems for the initial and final states were constructed by \textit{Kimocs}. These systems had the dimensions presented in table \ref{table:dimensions}. The bottom layer atoms were fixed. The initial and final states were relaxed using the conjugate gradient method. By linear interpolation between the atom positions of the initial and final states, 40 images were created. These images were used for the NEB calculation with PARCAS. Periodic boundaries in lateral $x$ and $y$ directions along the surface were used. No temperature nor pressure controls were used.

For the NEB calculations and MD simulations, we chose the interatomic potential based on the Corrected Effective Medium Theory (CEM), developed by M. S. Stave \textit{et al}.\;\cite{stave1990corrected}. The potential describes well the properties of Cu surfaces \cite{sinnott1991corrected}. For instance, the potential predicts the $\{111\}$ surface to be the most stable with the surface energy 1.76 J/m$^2$, while the surface energies of $\{100\}$ and $\{110\}$ are 1.91 J/m${^2}$, and 2.08 J/m${^2}$, respectively \cite{sinnott1991corrected}. For comparison, DFT gives 1.952 J/m$^{2}$ for $\{111\}$, 2.166 J/m$^{2}$ for $\{100\}$, and 2.237 J/m$^{2}$ for the $\{110\}$ surface  \cite{vitos1998surface}. The experimental value of the surface energy for the $\{111\}$ surface was reported to be $\sim$1.8 J/m$^{2}$ \cite{tyson1977surface,de1988cohesion}. As one can see, the surface energies given by the CEM potential are in good agreement with both DFT calculations and experiments.
\begin{table}
\centering
 \caption{The dimensions used for the simulated systems with different surfaces and the number of atomic monolayers (ML) used for the substrate. The $z$ direction is normal to the surface. }
\label{table:dimensions}
\begin{tabular*}{\columnwidth}{@{\extracolsep{\fill}} l l l l l l l}
 \toprule
 Surface	& $x$ [nm]	& $y$ [nm] & $z$ [nm]    & Substrate [ML]  		\\
 \midrule
$\{100\}$		& 7.4     & 7.4     & 14.8 & 12\\ 
$\{110\}$		& 12.8    & 9.0     & 5.2  & 18\\ 
$\{111\}$		& 9.0     & 5.2     & 12.8 & 30\\ 
\bottomrule
\end{tabular*}
\end{table}

\subsection{Molecular Dynamics simulations}\label{sec:MD}

We have also perform benchmarking MD simulations by using the PARCAS code  \cite{nordlund1998defect,ghaly1999molecular,nordlund1995molecular} with neither pressure nor temperature control. The dimensions of the simulation cells and the number of atoms were selected to match the ones used in \textit{Kimocs}. Periodic boundary conditions where used in $x$ and $y$ directions, but not in $z$. The bottom layer atoms were fixed. The CEM potential for Cu \cite{stave1990corrected}, which was used to calculate the barriers, was used also in the MD simulations. The simulations were performed with the time step 4.06 fs until the surface nanotip had flattened down to half of its initial height; that is, the simulations stopped when no atoms were above a certain $z$ coordinate.

\section{Parameterization of the KMC model}\label{sec:parameterization}

The parameterization of the Cu material, which we used for our model, can be summarized as follows:
\begin{itemize}
 \item The migration energy barriers are calculated using the NEB method;
 \item Processes involving atoms in unstable initial or/and final positions are treated separately. These positions may only appear due to the adopted rigid lattice approach and are usually parts of multiple transitions.
 \item The attempt frequency $\nu$ is estimated from a fit to corresponding MD data
\end{itemize}
These points will be described in detail in the following three subsections.

\subsection{Migration barrier calculations}\label{sec:migbarriers}

The evolution of the simulated system in \textit{Kimocs} is driven by diffusion jumps of atoms from occupied (initial) sites to unoccupied (final) ones in the 1nn vicinity of the former. Each jump in the system is associated with an energy barrier, $E_m(a,b,c,d)$, where $a$, $b$, $c$, and $d$ are the number of 1nn and 2nn atoms in the initial and final states (for details, see section \ref{sec:Kimocs}). The values of $E_m(a,b,c,d)$ are calculated using the NEB method implemented within the MD code PARCAS \cite{nordlund1998defect,ghaly1999molecular,nordlund1995molecular}. 
 The intrinsic feature of the adopted model assigns the value of $E_m(a,b,c,d)$, calculated for the randomly selected positional configuration described by the $a$, $b$, $c$, and $d$ indices, to all processes with the same set of these values. Although these identically interpreted transitions, which differ from one another by configurational arrangement only, may have slightly different energy barriers, we currently neglect these differences for the sake of computational efficiency. An exact description of the 1nn atom jumps with all positional permutations included would require about $10^7$ barriers to be calculated. However, ignoring the permutations and classifying all the transitions only by the  $(a,b,c,d)$ indices, the number of required calculations is reduced by several orders of magnitude.

We used the following algorithm to calculate the migration barriers. For representative positional permutations of a certain $(a,b,c,d)$ transition, we recorded possible situations which were probable to occur in the atomic system of interest by running \textit{Kimocs} for populations of randomly distributed adatoms on $\{100\}$, $\{110\}$, or $\{111\}$ Cu surfaces. We also used a bulk system with a random distribution of vacancies to characterize the processes with a high number of 1nn and 2nn atoms.

The values of the migration barriers were calculated as follows. Firstly, the initial and final states of the atomic systems for a given transition (an atom jump) were relaxed using the conjugate gradient method. After that the minimum energy path for the jumping atom to perform the transition between the initial and final states was found using the NEB method. The energy barrier, $E_m(a,b,c,d)$, was defined as the difference between the saddle point (the potential energy maximum) of the minimum energy path and the potential energy of the initial state. These energy barrier values were then added to the database to be used in \textit{Kimocs} for the actual simulations.

\begin{figure}
\centering
\subfigure[]{\includegraphics[width=0.4\columnwidth]{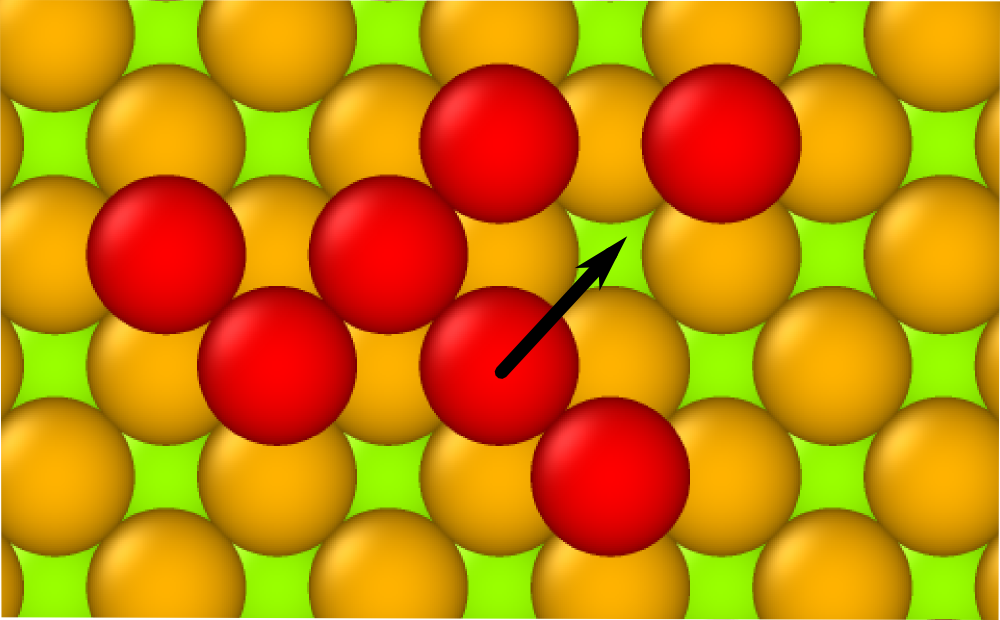}}
\subfigure[]{\includegraphics[width=0.4\columnwidth]{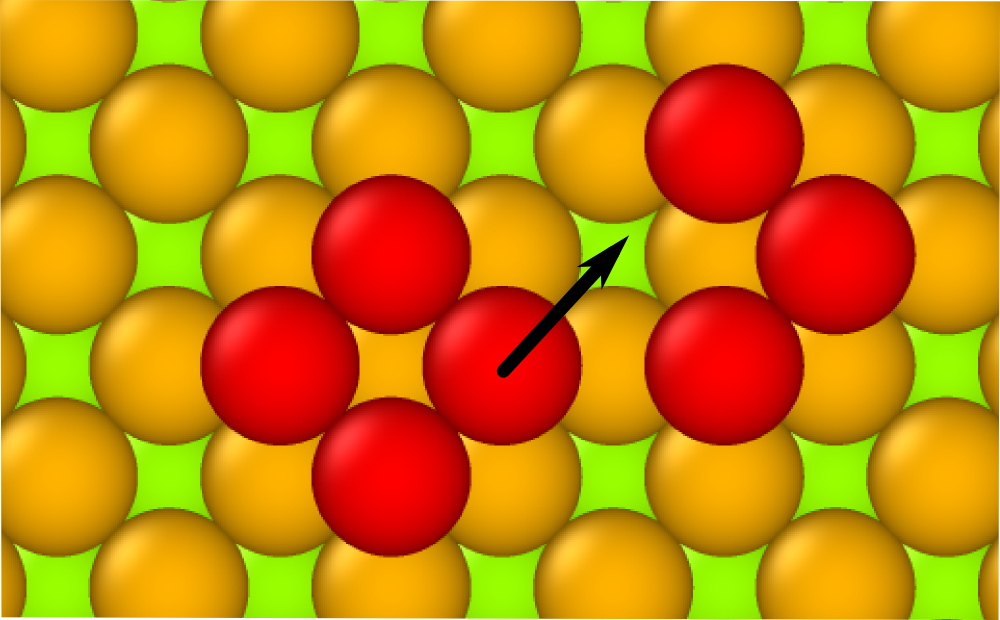}}
\caption{(Colour online) Two different permutations of the $(6,3,7,3)$ atom jump process.}
\label{fig:barrier_sensitivity}
\end{figure}

We estimated the uncertainty of the barrier values obtained by using the proposed $(a,b,c,d)$ characterization. We calculated the energy barriers for different jump processes identified with the same $(a,b,c,d)$ values, as is the case for, e.g., the processes shown in figure \ref{fig:barrier_sensitivity}. On average, five different permutations were calculated for 196 different $(a,b,c,d)$ transitions on a perfect $\{100\}$ surface with randomly distributed adatoms. The average standard deviation of the energy barriers for these 196 configurations was found to be 0.13 eV or 14.8\%, which gives an indication of the precision of our energy barrier characterization and subsequently of our KMC model in general. 

In some NEB relaxation simulations, atoms may diverge from the intended transition. In this kind of situation, the obtained value of the energy barrier would not correspond to the sought $(a,b,c,d)$ transition, but another process which is described by the $(a,b,c,d)$ indices corresponding to the states where the jumping atom found itself after the relaxations. During the calculations we also encountered situations in which the atoms were relaxing into different final positions than the expected ones. In these situations we ignored the exact position while calculating the barriers since only the values of the potential energies in the saddle point and the initial state were important. Since the atom relaxed into a different final state in the NEB calculations, it may be rather unphysical to force this atom into a position that is energetically unstable. This, however, will be taken into account during the next KMC step as the barrier to the stable position will have a near-zero value (section \ref{sec:transitions}), which can be interpreted as a spontaneous transition.

In principle, the small nanosize surface features must be studied by taking into consideration the finite-size effect due to the large surface areas with respect to the atoms in the bulk of the nanotips, which may affect dramatically the material properties. For instance, the melting point of thin nanowires has been shown to drop rapidly when the diameter of nanowire was below 20 nm \cite{granberg2014investigation}. In the same reference it was shown that 1 nm thick nanowires may melt already at room temperature. Although this process is well captured by the used CEM potential, this effect is currently not introduced in our model. The problem with unstable neighbour atoms was accentuated for migration barrier calculations on small nanotips, as the whole structure was much less stable during the relaxation process. Therefore, more robust systems such as plane surfaces or voids in the bulk, were preferred for the barrier calculations.

\subsection{Transitions involving atoms in unstable positions}\label{sec:transitions}

We noticed that atoms with no more than three nearest neighbours ($a \leq 3$) have near zero migration barriers to perform jumps to any 1nn lattice position. These processes take place instantaneously and are thus spontaneous. In the terminology of \textit{Kimocs}, these atoms are said to be in unstable positions.
To avoid the zero barrier problem, we apply a very small migration energy barrier to imitate the spontaneous jumps of unstable atoms:
\begin{equation}\label{eq:unstable_atoms}
 E_m(a,b,c,d) =  \epsilon a + \delta b + \epsilon c^{-1} + \delta d^{-1}
\end{equation}
where $a\leq3$, $c>0$, and $d>0$. For processes with $d = 0$, $E_m = 10^{-4}d^{-1}$ is used. $c>1$ is always true, since we count the jumping atom as 1nn of the final site. Although the differences are not expected to be large, this formula gives a priority to unstable atoms with less neighbours to jump before the unstable atoms with more neighbours. We assume that an atom jump to a position with a higher number of neighbours is more favourable: higher $a$ and $b$ thus raise the barrier, whereas higher $c$ and $d$ lowers it. We also take into account that 1nn atoms contribute stronger in the value of migration barriers compared to 2nn atoms by setting the corresponding contributions $\epsilon=10^{-3}$\;eV and $\delta=10^{-4}$\;eV for the 1nn and 2nn atoms, respectively, in order to obtain the right trend. Using these parameters, the maximum migration energy for an unstable atom is $E_m(3,6,1,1) = 0.0047$\;eV, which is insignificant compared to even thermal energies ($>$0.025\;eV). This is why the processes described by these barriers will appear spontaneous in the simulations. The low energy barriers ensure that the time increments, calculated by \ref{eq:residence_time}, are not overestimated.

\subsection{The attempt frequency}\label{sec:nu_fitting}

Another important parameter for KMC simulations is the attempt frequency $\nu$ for a jump to happen (\ref{eq:arrhenius}). This parameter affects the time predicted for a studied process to be completed. Since all processes in our model are jumps by atoms to 1nn lattice sites, the attempt frequencies can be reasonably assumed to be approximately the same for all transitions. It is also frequently assumed to be of the same order of magnitude as the Debye frequency (for Cu, $\nu=4.5\cdot10^{13}$\;s$^{-1}$)  \cite{hook2013solid,soisson2007cu,vincent2008precipitation,castin2011modeling,jansson2013simulation}. In our model we fitted the value $\nu$ to the MD simulations (for simulation details, see section \ref{sec:MD}), comparing the flattening time of a surface nanotip obtained by both methods. The fitting procedure can be described as follows.

\begin{figure}
    \centering
    \includegraphics[width=\columnwidth]{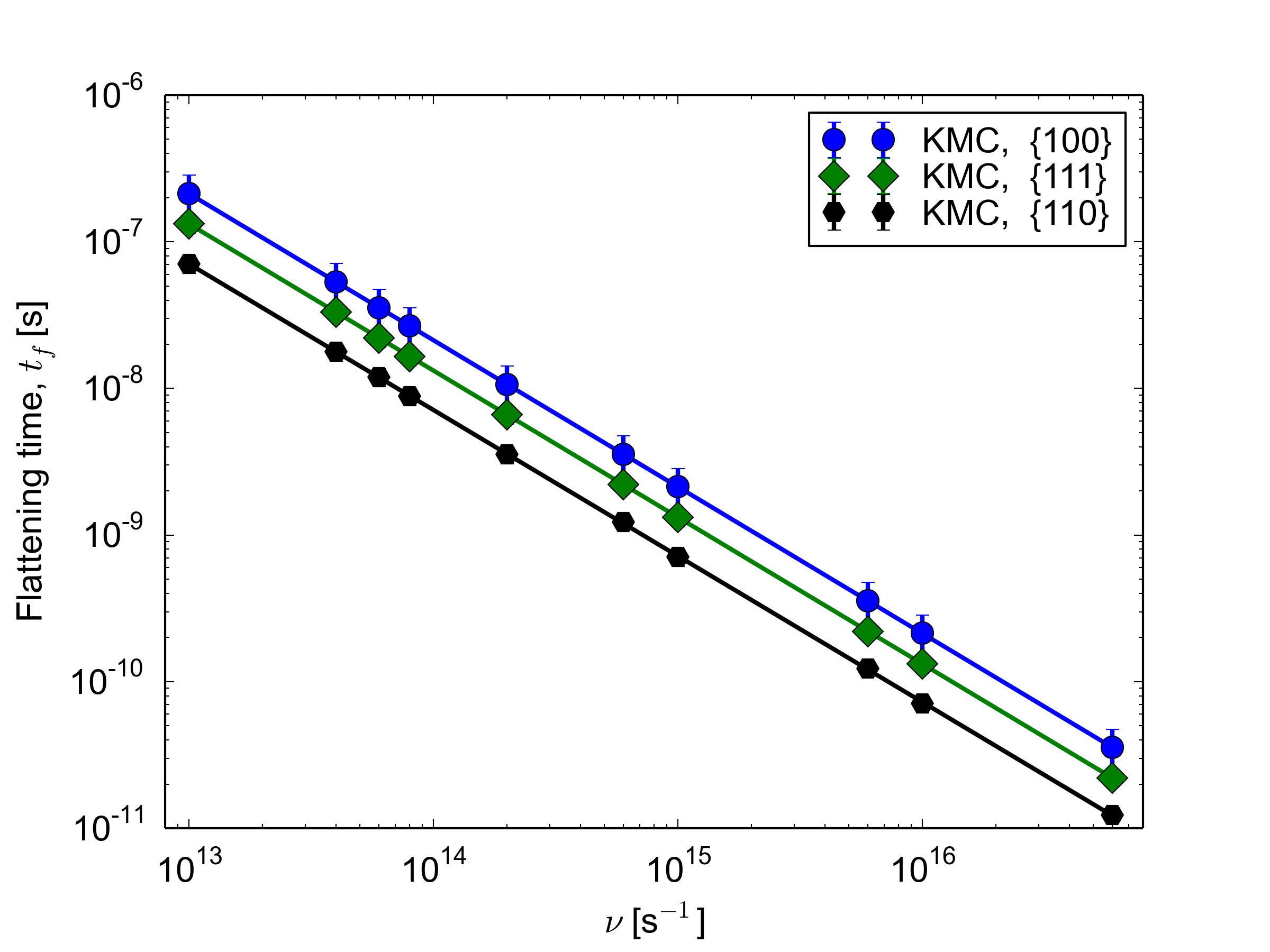}
   \caption{(Colour online) The flattening time $t_f$ for the simulated 12 ML nanotip as a function of the attempt frequency $\nu$ used in \textit{Kimocs}. The flattening time is calculated as the time elapsed for the nanotip to flatten down in height from 12 to 6 ML. Three different surfaces (lattice orientations) were used: \{100\}, \{110\} and \{111\}.}
    \label{fig:attempt_freq_fit}
   \end{figure}

A small cuboid nanotip with 576 atoms and a height of 12 monolayers (ML) were placed on three different Cu surfaces: \{110\}, \{111\}, and \{100\}, respectively. The system dimensions are listed in table \ref{table:dimensions_attempt}. Periodic boundaries were used in the lateral $x$ and $y$ directions. The bottom layer of atoms was fixed and monitored throughout the simulations not to interact with the jumping atoms of the surface. The time elapsed for the cuboid nanotip to flatten down in height from 12 to 6 ML at 1000 K with different attempt frequencies was recorded. The statistical uncertainty was taken into account by using ten different seed numbers for every value of the attempt frequency.  The obtained flattening times are plotted in figure \ref{fig:attempt_freq_fit}. The KMC values for the flattening time $t_f$ could be fitted by the function
\begin{equation}\label{eq:nu_fit}
 t_f = \nu^{-1} \rme^c, 
\end{equation}
where the constant $c$ can be interpreted as $c=E_m/(k_B T)$, where $E_m$ is the average atom migration barrier at the temperature $T$ and $k_B$ is the Boltzmann constant. Here, $E_m$ is 1.156\;eV for \{110\}, 1.216\;eV for \{111\}, and 1.256\;eV for the \{100\} case, respectively. These values do indeed correspond well to the barriers used in our KMC model. Comparing with the MD values (the tabulated values and discussions can be found in section \ref{sec:benchmarking}), gives the attempt frequency values $\nu$ for the different surfaces as $7\cdot10^{13}$\;s$^{-1}$ (\{110\}), $2\cdot10^{14}$\;s$^{-1}$ (\{111\}), and $1\cdot10^{15}$\;s$^{-1}$ (\{100\}), respectively. We chose $\nu = 7\cdot 10^{13}$\;s$^{-1}$, as it is closest to the Debye frequency value of Cu.
\begin{table}
  \centering
   \caption{The dimensions of different simulation cells with three different surface directions used for determining the attempt frequency in MD and KMC. The number of atom monolayers (ML) indicates the thickness of the substrate, on which the nanotip was placed. The $z$ dimension is normal to the surface.}
\label{table:dimensions_attempt}
\begin{tabular*}{\columnwidth}{@{\extracolsep{\fill}} l l l l l}
\toprule
 Surface	& $x$ [nm]	& $y$ [nm] & $z$ [nm]    & Substrate [ML]  		\\
 \midrule
$\{100\}$		& 7.4     & 7.4     & 14.8 & 12\\
$\{110\}$		& 12.8    & 9.0     & 5.2  & 18\\
$\{111\}$		& 9.0     & 7.8     & 12.8 & 30\\
\bottomrule
\end{tabular*}
\end{table}

\section{Results}\label{sec:results}

\subsection{Validation of the model}\label{sec:validation}

\subsubsection{Flattening time of a nanotip on different surfaces.}\label{sec:benchmarking}

\begin{figure*}
\subfigure[]{
  \includegraphics[width=0.3\textwidth]{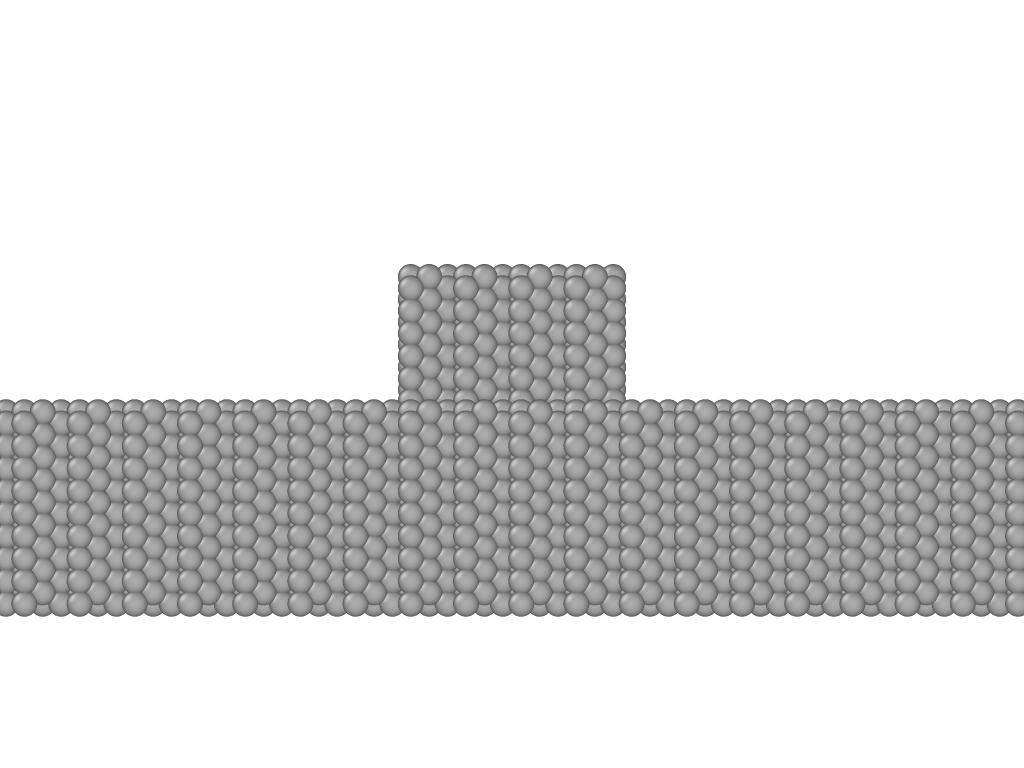} 
  }
\subfigure[]{
  \includegraphics[width=0.3\textwidth]{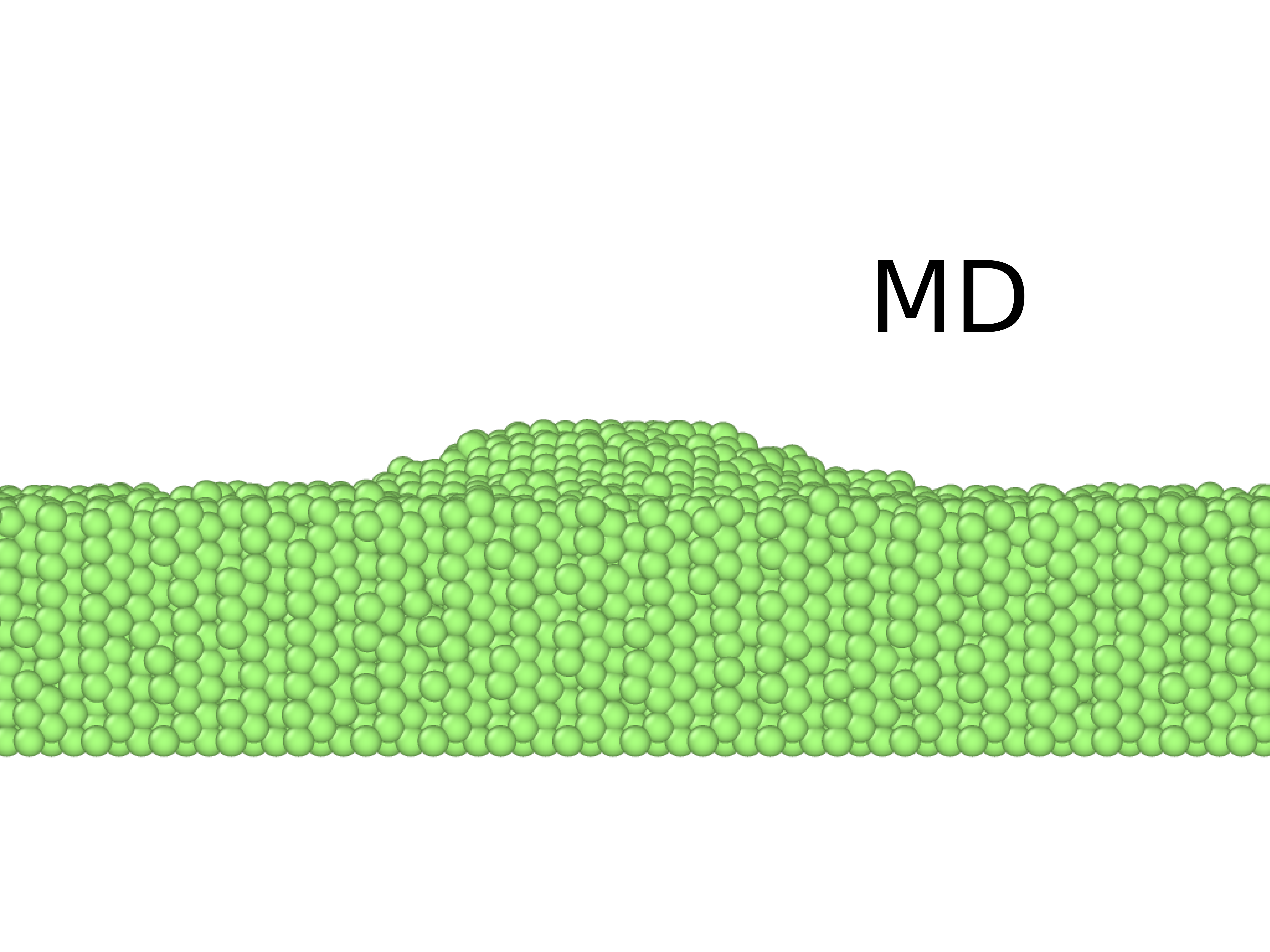} 
}
\subfigure[]{
  \includegraphics[width=0.3\textwidth]{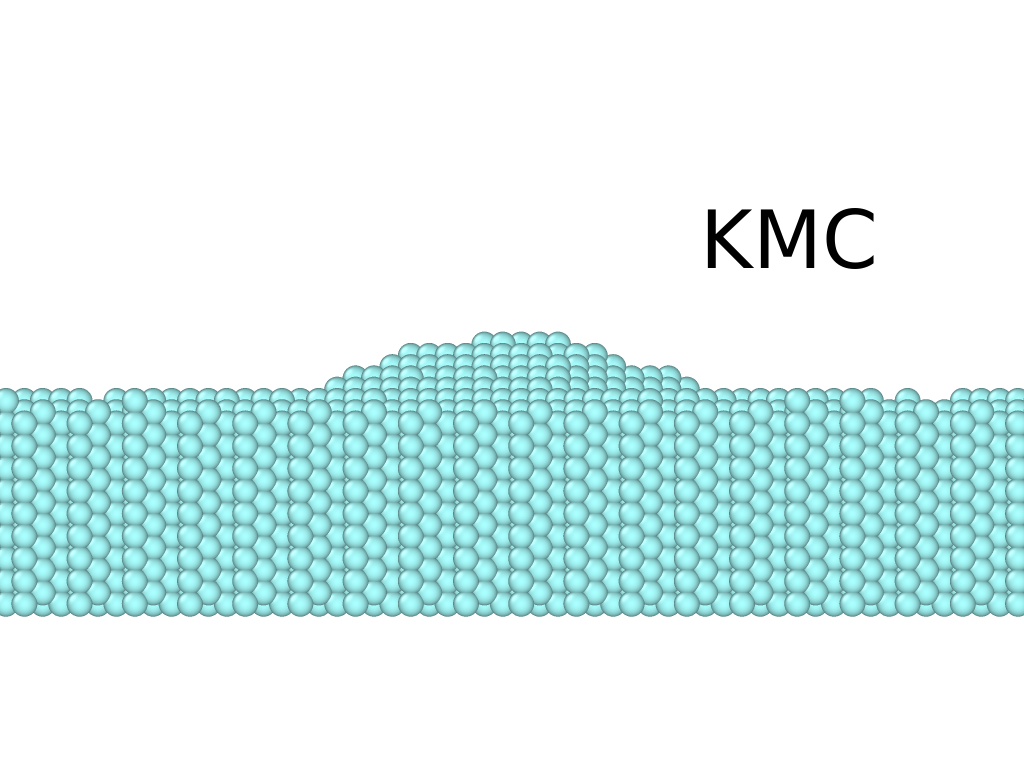} 
}
\caption{(Colour online) Final snapshots of MD (b) and KMC (c) simulations of flattening of a 12\;ML nanotip, initially in the shape of a cuboid (a). In both methods the simulations were stopped when the nanotip height was reduced by half. The surface evolutions predicted by both methods are identical (see comparative animation in the Supplementary Material).}
\label{fig:tip_md_kmc}
\end{figure*}

Using the parameterization described in section \ref{sec:parameterization}, we carried out a series of \textit{Kimocs} simulations to validate our model against the  corresponding MD results. At first we analysed the flattening process of 12 ML high cuboid surface nanotips with respect to the crystallographic orientation of the surface. The details of the simulation setups are the same as in section \ref{sec:nu_fitting}. The temperature was set to 1000 K and the simulations were stopped when the height of the cuboid nanotips had decreased to 6 ML (half of their original size). The simulations were repeated with ten different seeds and the average flattening time was recorded for each surface, as seen in table \ref{table:R20150209}.
In order to validate the KMC model, the results have been compared with MD simulations, as also shown in table \ref{table:R20150209} and figure \ref{fig:tip_md_kmc}.
The results of our KMC simulations agree with the MD results. We also note that the KMC simulations were two orders of magnitude faster computationally than the MD simulations.
\begin{table}
  \centering
   \caption{Flattening of cuboid nanotips, 12 ML high with 576 atoms, on different surfaces at 1000 K; comparing MD and KMC. The simulations were stopped when the nanotips had flatten below half their original height (6 ML).} 
\label{table:R20150209}
\begin{tabular*}{\columnwidth}{@{\extracolsep{\fill}} l l l}
\toprule
 Surface	& MD [ns]	& KMC [ns] 		\\
 \midrule
$\{110\}$	& $9.29\pm1.44$ & $9.25 \pm 1.10$	\\
$\{111\}$	& $6.01\pm1.48$ & $18.8 \pm 0.96$	\\
$\{100\}$	& $1.62\pm0.60$ & $31.0 \pm 6.61$	\\
\bottomrule
\end{tabular*}
\end{table}

The results presented in table \ref{table:R20150209} show a very good agreement between the results obtained with KMC and MD for the nanotips built on a \{110\} surface as they are the most stable ones. The nanotips built on the \{111\} and on \{100\} surfaces compare worse, which can be explained by the shape transformations of small scale (diameter $\sim$2\;nm) nanotips. These transformations are seen in MD simulations \cite{liang2005shape,park2005shape}, but not accessible by KMC due to the assumed rigid lattice of the simulated material. However, overall the agreement is fairly good and does not exceed the factor 20, taking into account the different nature of the two simulation techniques (no relaxation of the lattice is taken into account in KMC, also the barriers for other than the \{111\} surfaces can be overestimated by NEB)

\subsubsection{Thermal behaviour of surface nanotips.}\label{sec:T_dependence}

Analysis of the flattening time over the range of temperatures will enable the prediction of the stability of the small size nanotips at much lower temperatures (room and below), which are difficult to access even by KMC methods. For this, we repeated the same KMC simulations as in section \ref{sec:benchmarking} with temperatures ranging from 500 to 1200\;K. For comparison, MD simulations of the same systems were performed only for a higher temperature range of 850\;K to 1200\;K. The KMC data points from 900\;K and higher were averaged over 10 runs per temperature, whereas the data points for lower temperatures were performed only once. The simulations were stopped when the cuboid had been flatten down to below half its original height (6\;ML). The flattening time $t_f$ was found to follow Arrhenius-like behaviour:
\begin{equation}\label{eq:flattening}
t_f = t_0 \exp \left( \frac{E_a}{k_BT} \right),
\end{equation}
where the prefactor is $t_0 = 2.34\cdot10^{-12}$\;s, the activation energy is $E_a = 0.72$\;eV, and $k_B$ is the Boltzmann constant. These can be compared to the values fitted to the MD data: $t_0 = 7.33\cdot10^{-14}$\;s and $E_f = 1.00$\;eV. Both sets of $t_0$ and $E_a$ compare well to the average migration barrier and inverse of the attempt frequency of the diffusion jumps (1.04\;eV and $1.4\cdot10^{-14}$\;s, respectively). The difference may be related to the limitation of the NEB method, which predicts the most relaxed pathways for atomic transitions. In MD, some transitions with the barriers higher than predicted by NEB may naturally occur, especially at high temperatures.

\begin{figure}
 \includegraphics[width=\columnwidth]{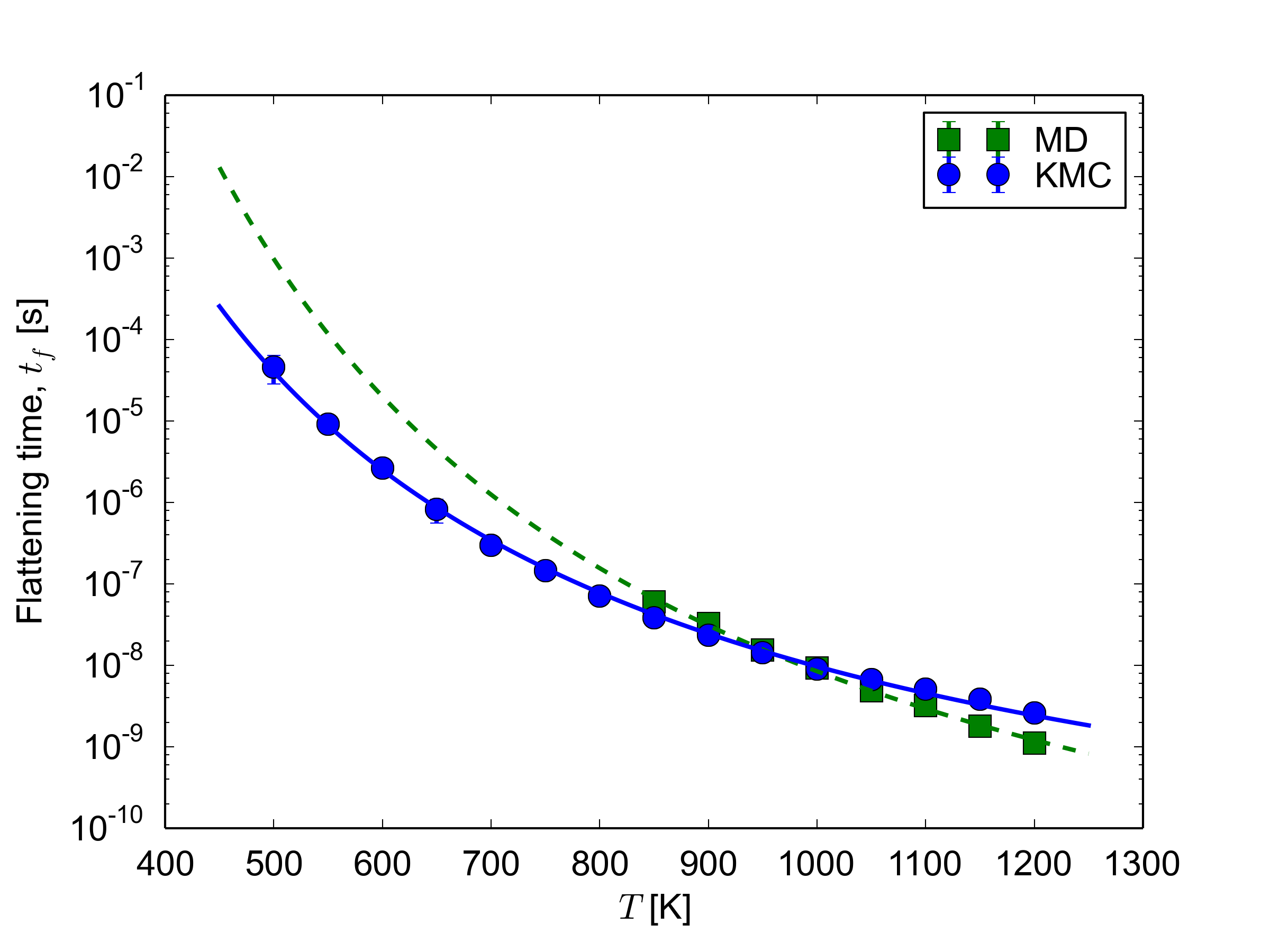}
 \caption{(Colour online) Flattening time with KMC and MD for a cuboid nanotip with a height of 12 ML (1.6\;nm) and 576 atoms on a $\{110\}$ surface as a function of temperature. The lines are the fits of \ref{eq:flattening} to the KMC (solid) and MD (dashed) results.}
 \label{fig:T_dependence}
\end{figure}

The results are plotted in figure \ref{fig:T_dependence}. At temperatures between 800 and 1100 we see a very good agreement of the KMC and MD results. The temperature dependence for the flattening time shows very similar trends for both methods, with the KMC data showing slightly weaker dependence on the temperature than the MD. At the low temperatures (below 850 K), comparison is not possible, since MD is too slow to produce any sensible data in this regime. The dashed line is an extrapolation obtained by fitting \ref{eq:flattening}, to the MD data (filled squares).

\subsubsection{Formation of adatom islands.}

We have also analysed the migration of single adatoms on the surface to assess whether the model can capture satisfactorily the surface diffusion on Cu surfaces. We studied the dynamics of adatom migration by distributing randomly 300 adatoms on an atomically smooth surface. Three different surfaces were considered: $\{100\}$, $\{110\}$, and $\{111\}$. The dimensions of the surfaces were $10\times 10$\;nm$^2$. We saw that adatom islands were formed in less than 1 ns on all three surfaces. Moreover, the bigger islands were growing on the expense of smaller ones according to the Ostwald ripening mechanism.
\begin{figure}
\centering
 \subfigure[$\{100\}$]{
  \includegraphics[width=0.6\columnwidth]{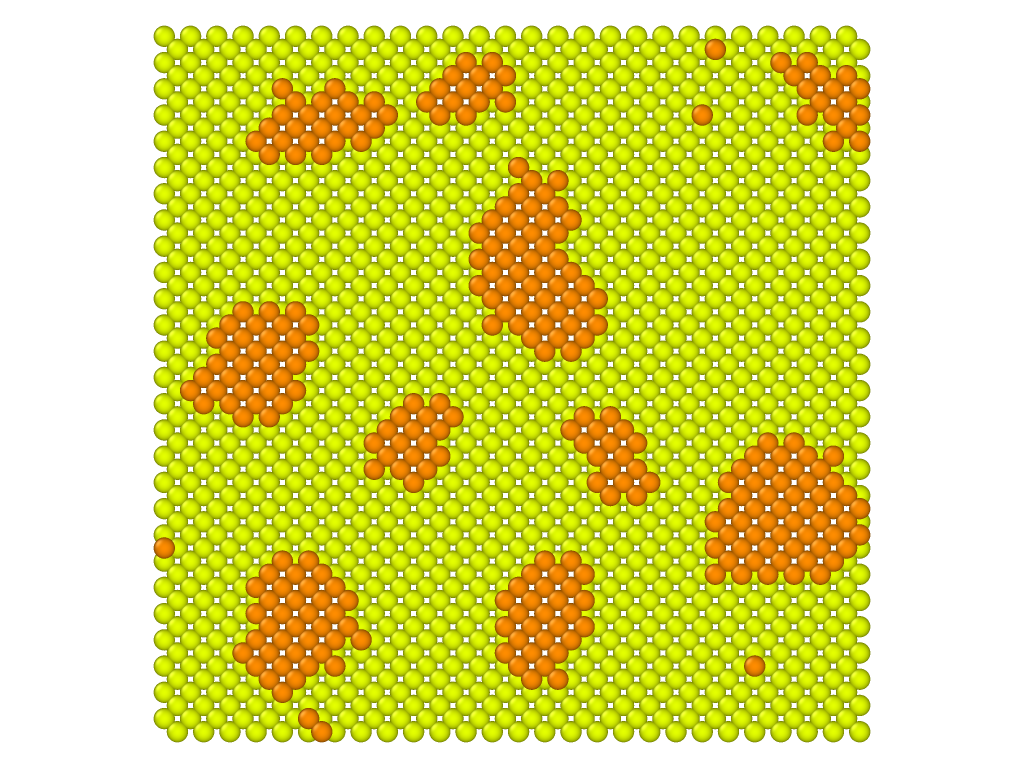}
  }
 \subfigure[$\{111\}$]{
 \includegraphics[width=0.6\columnwidth]{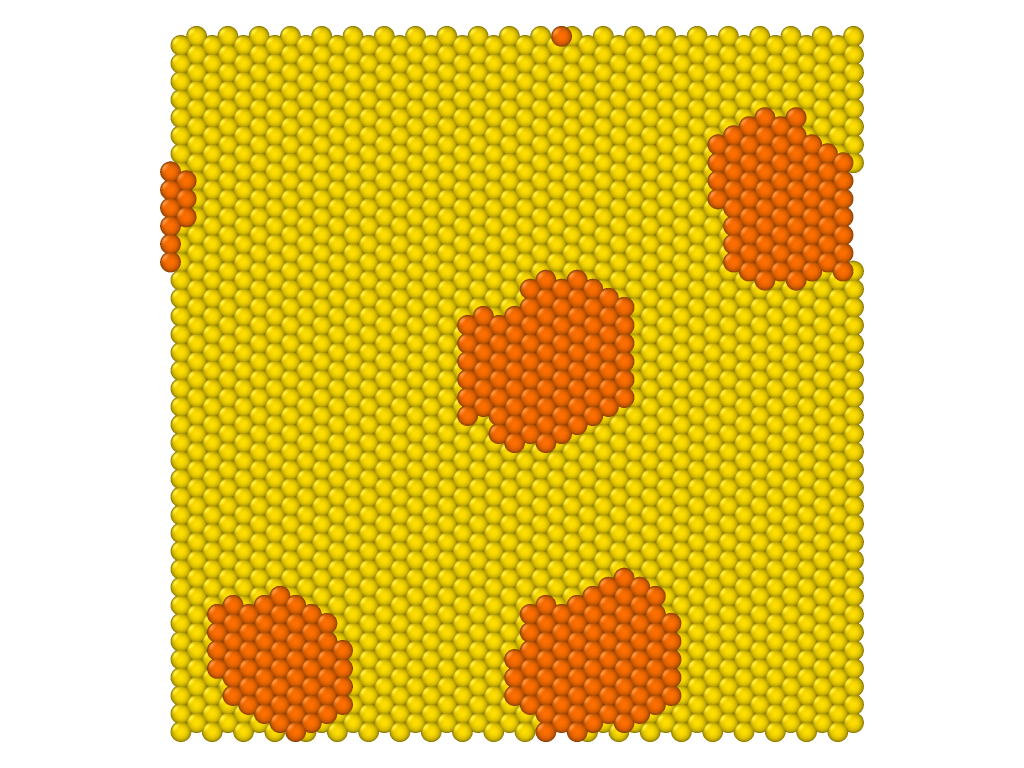}
  }
\subfigure[$\{110\}$]{
 \includegraphics[width=0.6\columnwidth]{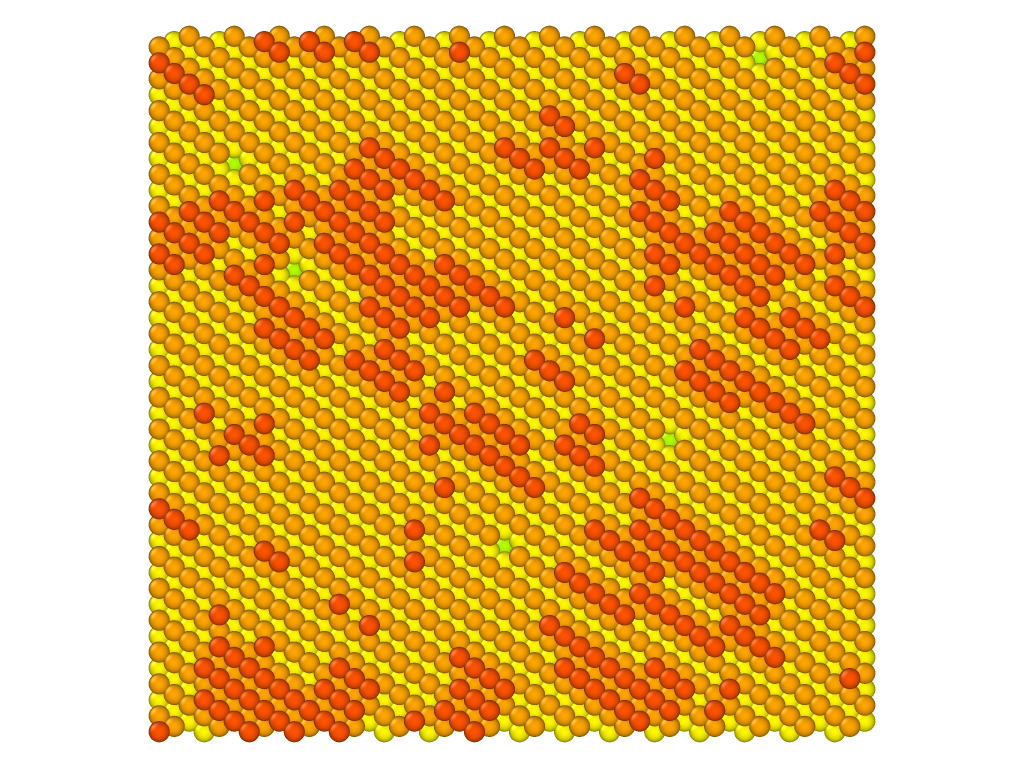}
 }
 \caption{(Colour online) Coalescence of adatoms resulting in nanoislands on the $\{100\}$, $\{111\}$, and $\{110\}$ surfaces.}
 \label{fig:adatom_islands}
\end{figure}

In figure \ref{fig:adatom_islands} we show examples of the nanoislands formed on all three different surfaces. As one can see the faster diffusion process on the \{111\} surface results in clearly separated and well defined big islands. The \{110\} surface exhibits the least pronounced structure as the preferential migration along $\langle$110$\rangle$ surface channels leads to the formation of elongated and less organized structures.

The adatom island dynamics produced by our KMC model is thus in good agreement with experiments \cite{hannon1997surface} and other KMC studies \cite{mottet1998monte,henkelman2001long,karim2006diffusion}.

\subsection{Stability of large nanotips}\label{sec:large_tips}

\subsubsection{The flattening process of large nanotips.}

Tall and sharp surface nanotips with high aspect ratios are believed to be responsible for the enhanced field emissions and, subsequently, the vacuum arcs observed in experiments with high electric fields \cite{navitski2013field,muranaka2011situ,nagaoka2001field}. The exact shape of these nanotips and how they are created is not known. Tips are not likely to be seen after vacuum arc events. It is also quite difficult to observe nanotips which may have grown under an electric field but have not yet caused a vacuum arc. No such evidences exist in the literature to our knowledge.
It suggests that the life time of such nanotips is too short to be observed with electron microscopes or other experimental techniques. Using our KMC model, we have simulated the flattening process of large narrow nanotips that may be considered as candidates for field-emitting nanotips. By estimating the flattening time of the nanotips at different temperatures, we have evaluated the stability and life time of field-emitting nanotips.

\begin{figure}
 \centering
 \includegraphics[width=\columnwidth]{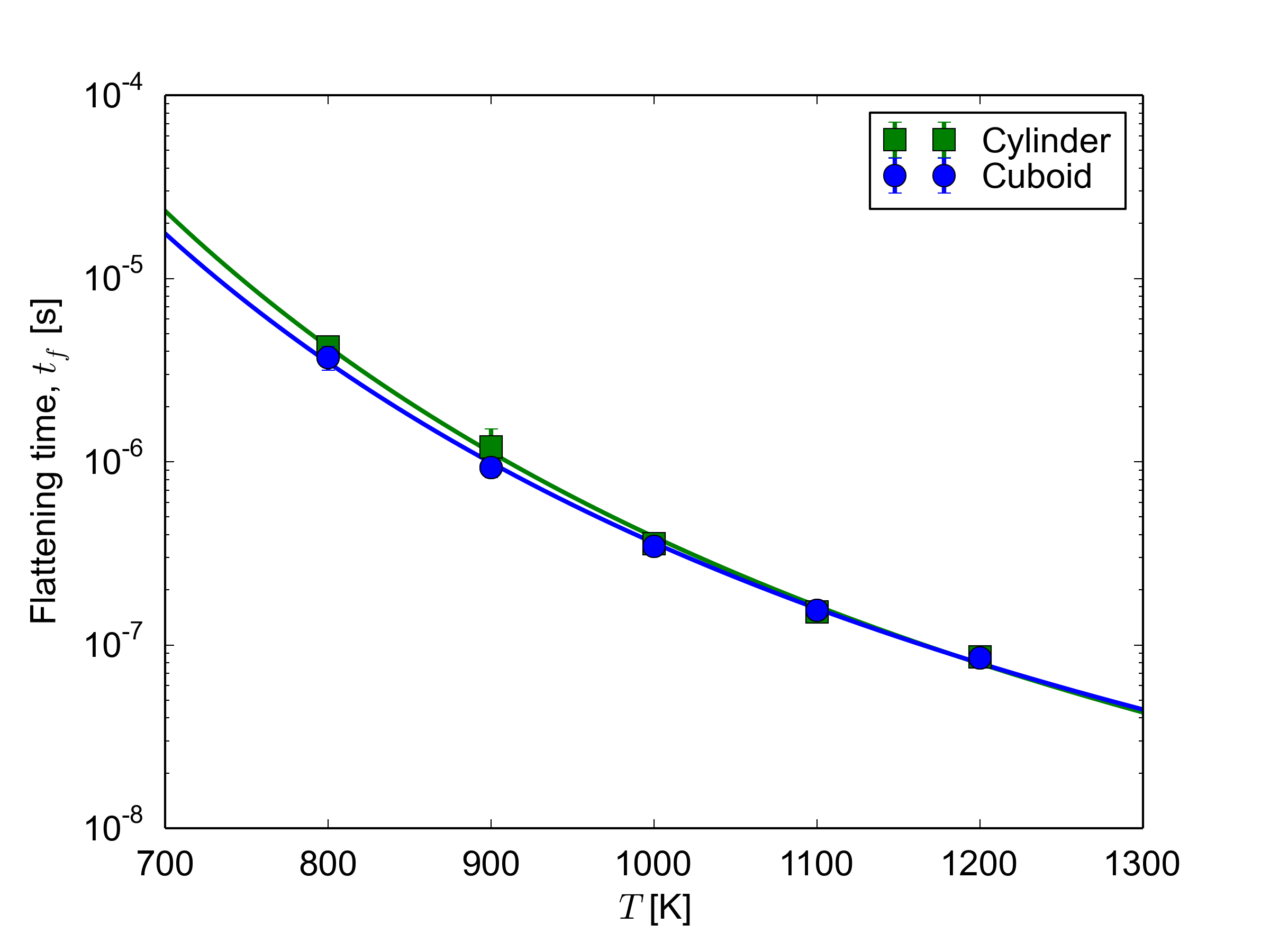}
 \caption{(Colour online) Flattening time for large 13\;nm high nanotips versus temperature. The simulations were stopped when the height of the nanotip was reduced to the half of its original size.}
 \label{fig:R20150327.png}
\end{figure}

In these simulations, a cuboid nanotip of 13\;nm in height and 2\;nm in width (aspect ratio $\sim$7) was constructed on a \{110\} surface. Consistently with our previous simulations, we continued the simulations of the tall nanotips until they shortened to the half of their original height, 7\;nm. We simulated the nanotips at different temperatures between 800\;K and 1200\;K. The results are shown in figure \ref{fig:R20150327.png}. We also simulated a nanotip of double height (26 nm; aspect ratio $\sim$14). At 1200\;K, the nanotip necked at $(164\pm20)$\;ns, which is about twice the flattening time of the 13\;nm tall nanotip at the same temperature: $(84.7\pm 5.5)$\;ns. 

\begin{figure*}
\centering
 \subfigure[$t = 0.0$]{
  \includegraphics[width=0.2\textwidth]{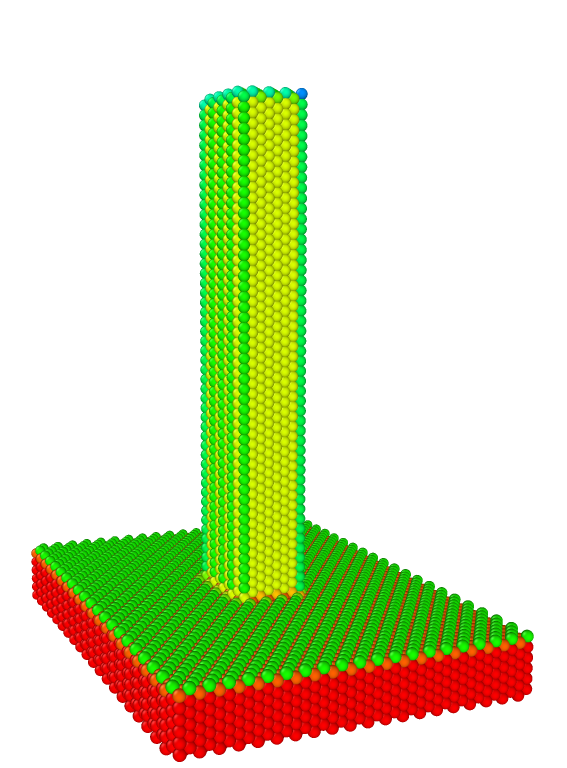}
 }
 \subfigure[$t = 0.2$\;µs]{
 \includegraphics[width=0.2\textwidth]{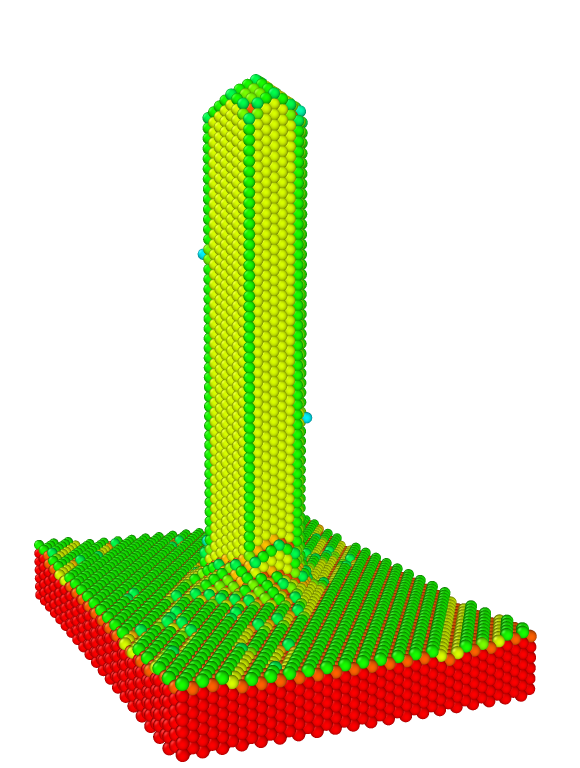} 
 }
 \subfigure[$t = 2.1$\;µs]{
 \includegraphics[width=0.2\textwidth]{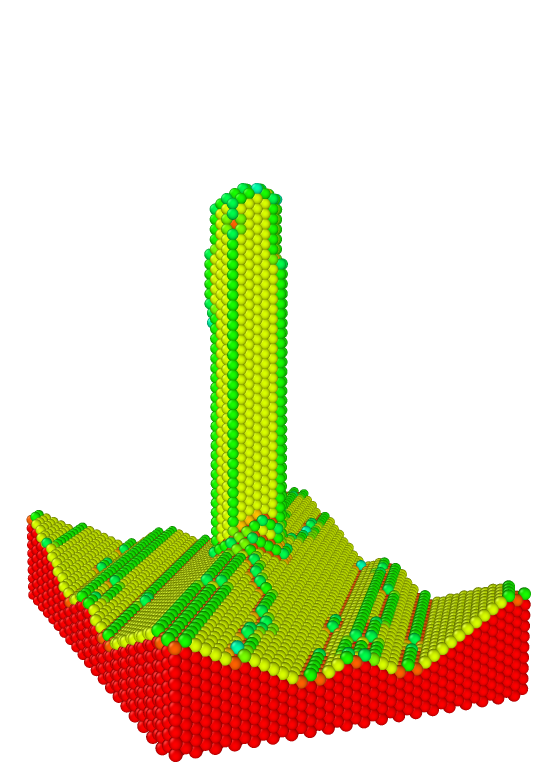}
 }
 \subfigure[$t = 6.3$\;µs]{
 \includegraphics[width=0.2\textwidth]{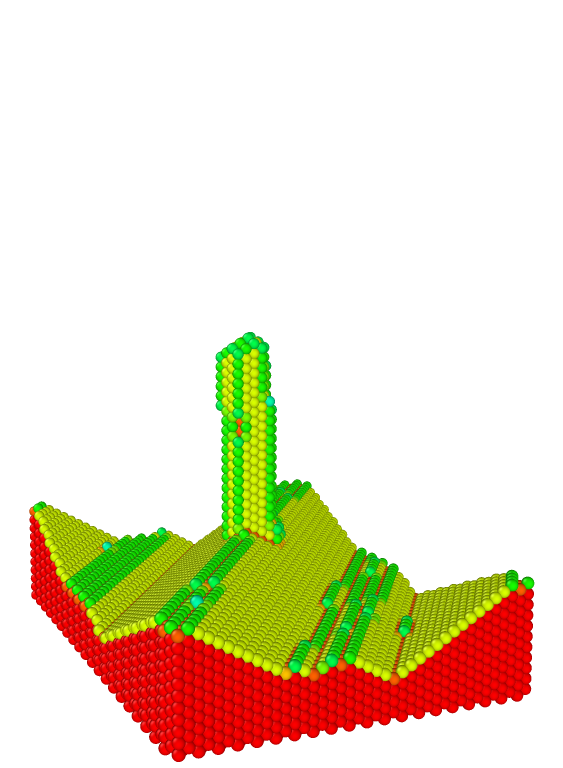}
 }
 \caption{(Colour online) The flattening of a 13\;nm high nanotip at 800\;K at (a) $t=0.0$, (b) after 0.2\;µs, (c) after 2.1\;µs, and (d) after 6.3\;µs, when it was reduced to half of its original height. The atoms are coloured according to their coordination number (amount of 1nn atoms) to highlight the faceting of the nanotip. The nanotip is initially a cuboid with a $\{110\}$ top surface and two $\{112\}$ and $\{111\}$ side surfaces (See figure \ref{fig:apexes} for a detailed view). Already after 0.2\;µs, the closed-packed $\{111\}$ facets (yellow) are dominating and the $\{112\}$ facets have disappeared. The ridges formed on the substrate surface is an artefact caused by the periodic boundary conditions.}
\label{fig:large_flattening}
\end{figure*}

\begin{figure}
 \includegraphics[width=\columnwidth]{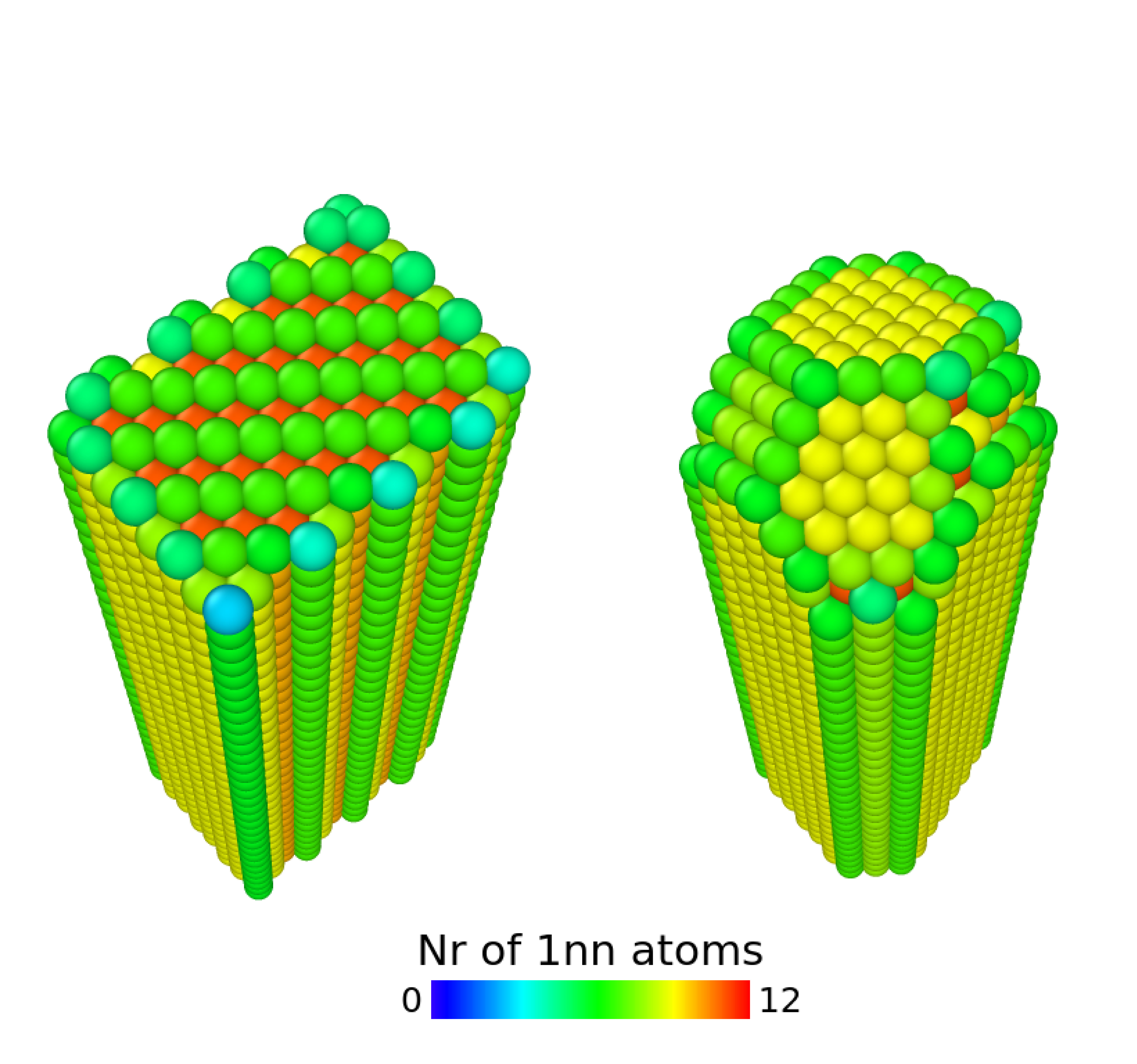}
 \caption{(Colour online) Detailed view of the apex of the 13\;nm nanotip in figure \ref{fig:large_flattening} at $t=$ 0.0 (left) and $t= 0.2$\;µs (right). The atoms are coloured according to their coordination numbers. The surface atoms become more bonded (due to the faceting) in the course of the simulation. }
 \label{fig:apexes}
 \end{figure}

In figure \ref{fig:large_flattening} we show the sequence of images (snapshots taken at 0.0\;µs, 0.2\;µs, 2.1\;µs, and 6.3\;µs of simulated time) of the flattening process of the 7\;nm nanotip at 1200 K. We see that the main mechanism for flattening is the diffusion of atoms down from the sides of the nanotip to the substrate. At first, the close-packed \{111\} facets are formed on the nanotip sides, as shown in figure \ref{fig:apexes}. This process stabilizes the nanotip as the atoms have more neighbours (bonds) within the plane, whereas the adatoms migrating on this plane have less bonds and thus migrate more easily. These facets then shrink layer by layer, as also these atoms diffuse towards the substrate. As in the case of the small nanotips in section \ref{sec:T_dependence}, the flattening time again follows an Arrhenius-like trend (\ref{eq:flattening}) with a prefactor $t_0 = 1.43\cdot10^{-11}$\;s and an activation energy of $E_a = 0.89$\;eV. The value of the activation barrier is greater in this case, which is explained by the strong faceting of the taller nanotips. The taller the nanotip the more atoms are bonded in the closed-packed facets, and hence a higher migration energy is required for the atoms to break out of the faceted plane and diffuse.

At 300 K, no significant change of the nanotip was observed after 10\;µs. Using \ref{eq:flattening}, the flattening time at 300\;K can be estimated to 3.1\;h. The faceting of the substrate is an artefact of the periodic boundary condition and is not the result of the current model.

Since we for simplicity have used cuboid nanotips for the model construction, we verified the shape independence by also repeating the simulations with a cylindrical nanotip. The cylindrical nanotip had the same height, 13\;nm, and number of atoms, 4800, as in the case of the cuboid nanotip. No significant difference in the flattening time was observed (figure \ref{fig:R20150327.png}).

\subsubsection{The pearling instability of nanotips and infinite nanowires.}

\begin{figure*}
 \centering
  \subfigure[]{
   \includegraphics[width=0.3\textwidth]{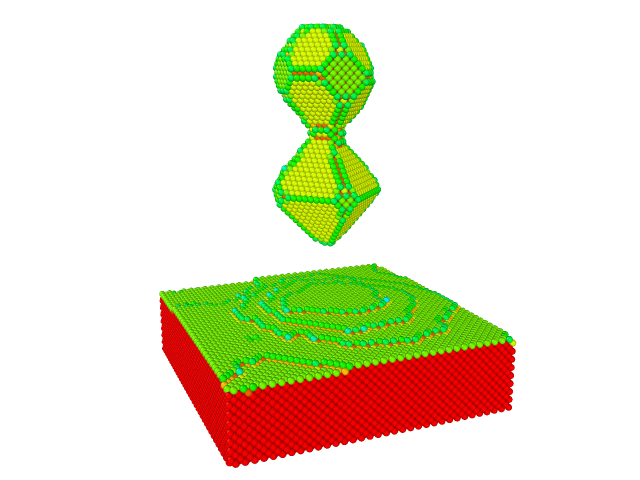}
  }
  \subfigure[]{
  \includegraphics[width=0.3\textwidth]{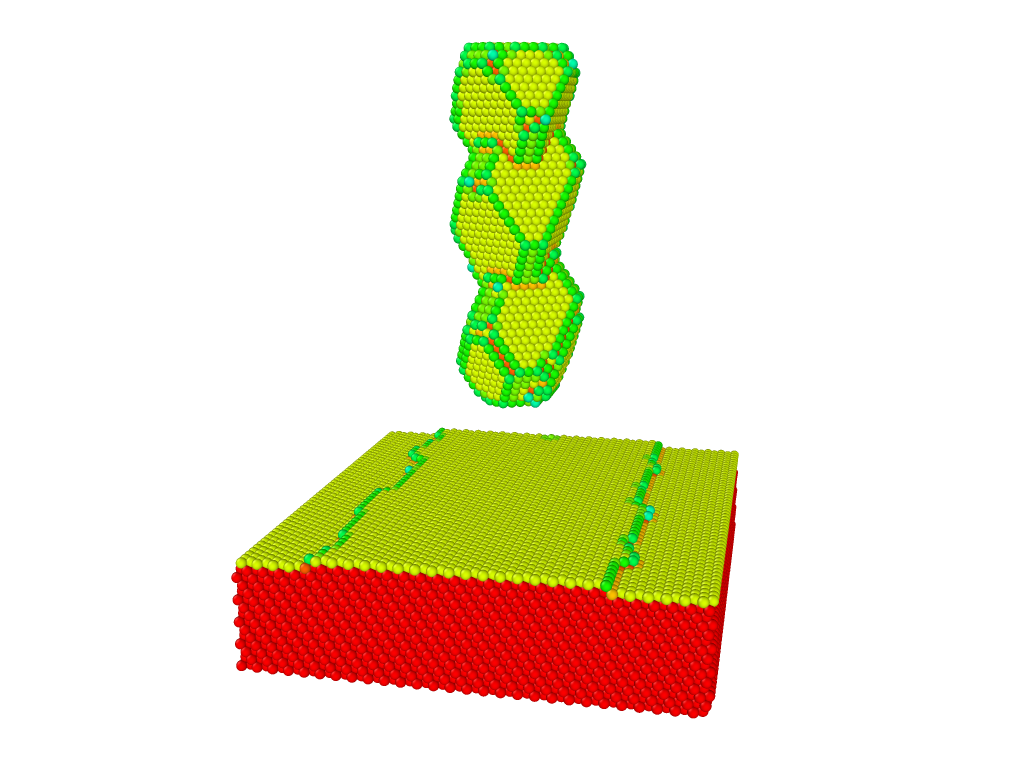}
  }
  \subfigure[]{
  \includegraphics[width=0.3\textwidth]{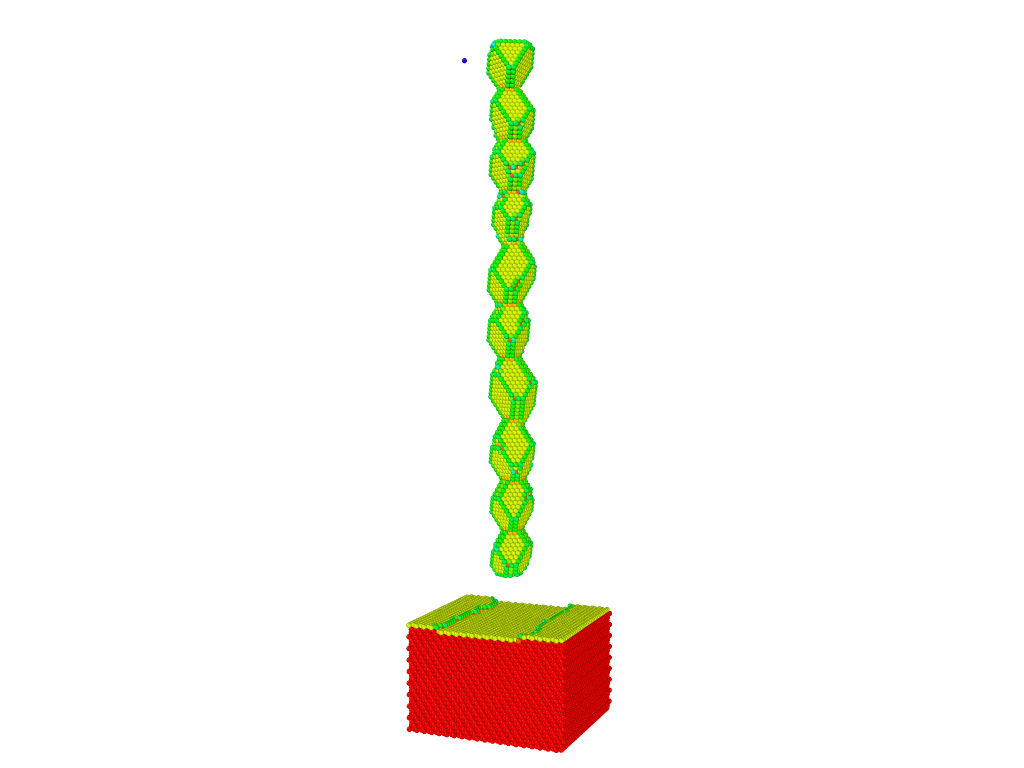}
  }
  \caption{(Colour online) The necking instability of a 13\;nm high nanotip at 1000\;K with $\langle 100 \rangle$ orientation after 400\;ns (a), the same for a $\langle 111 \rangle$ nanotip of identical height after 300\;ns (b). The $\langle 111 \rangle$ tend to neck near the substrate, whereas the broken-off structure obtains a regular faceted pattern that keeps stable also after 900\;ns, as shown in (c) for a 31\;nm high $\langle 111 \rangle$ nanotip. The atoms are coloured according to their coordination number in order to highlight the faceted structures.}
  \label{fig:pearling}
 \end{figure*}

During the simulations of tall nanotips on the \{100\} and \{111\} surfaces we found that they were unstable at 1000\;K with our KMC model. The nanotips were 13\;nm high with a diameter of 2.6\;nm. At 1000\;K, the nanotip with the major axis in the $\langle 100 \rangle$ direction exhibited Rayleigh necking and separated from the substrate after 70\;ns. After 400\;ns, the detached nanotip had changed into two barely attached polygon-shaped crystals [figure \ref{fig:pearling}(a)]. The nanotip oriented in the $\langle 111 \rangle$ direction also exhibited necking and detached from the surface after 100\;ns. The remainder of the suspended (not attached to the substrate) nanotip developed regular facets but was stable for at least another 300\;ns [figure \ref{fig:pearling}(b)]. A larger $\langle 111 \rangle$ nanotip of the height 32\;nm and the diameter 1.8\;nm also necked at the surface, but the upper part of the nanotip, which also developed regular facets as the smaller nanotip, remained stable for at least 900\;ns [figure \ref{fig:pearling}(c)]. This kind of Rayleigh instability is similar to the ``pearling instability'' effect that has been observed experimentally for Cu nanowires \cite{toimil2004fragmentation,zhao2006patterning,muller2002template}. 

\begin{figure*}
 \centering
  \subfigure[]{
   \includegraphics[width=0.3\textwidth]{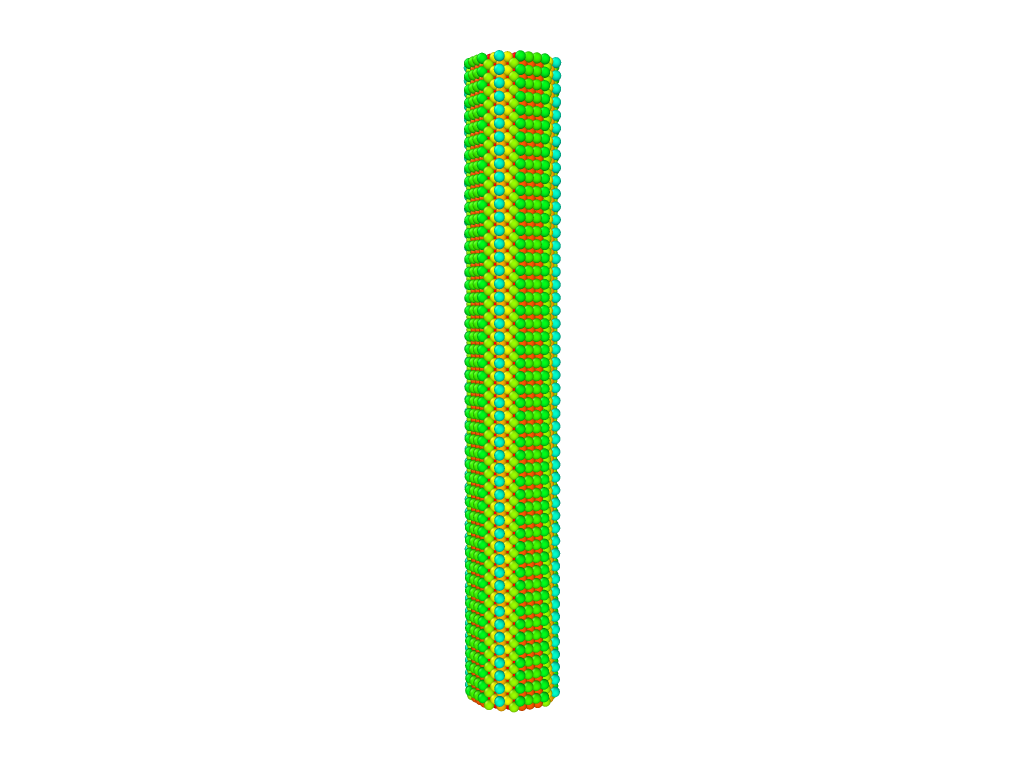}
  }
  \subfigure[]{
  \includegraphics[width=0.3\textwidth]{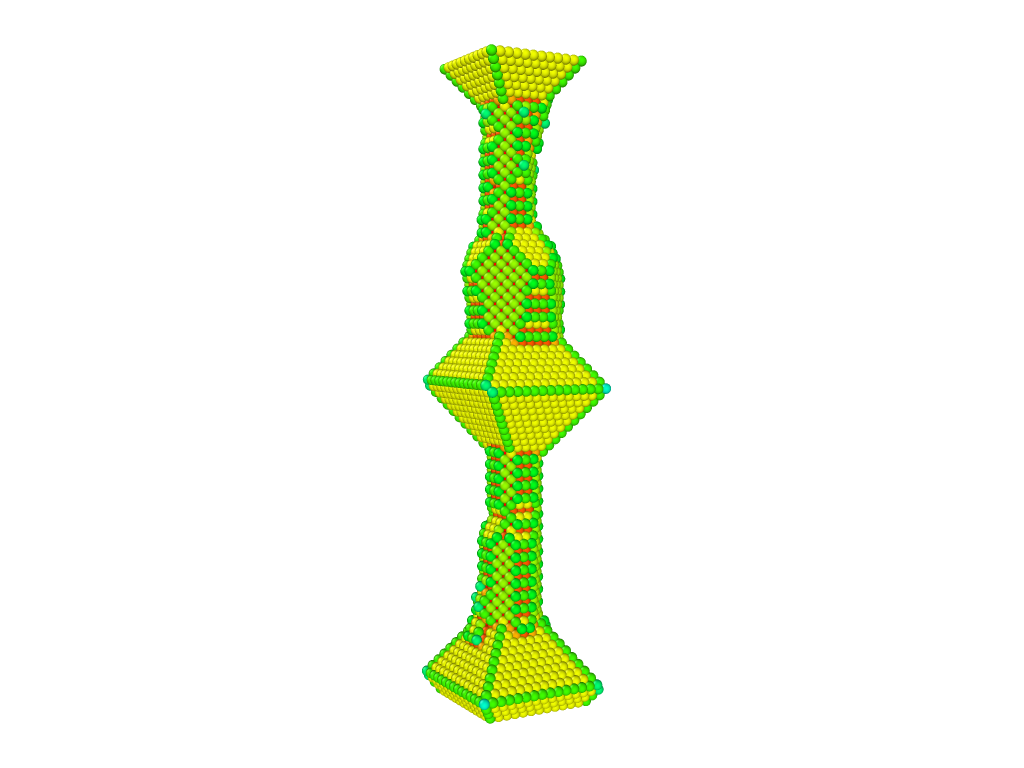}
  }
  \subfigure[]{
  \includegraphics[width=0.3\textwidth]{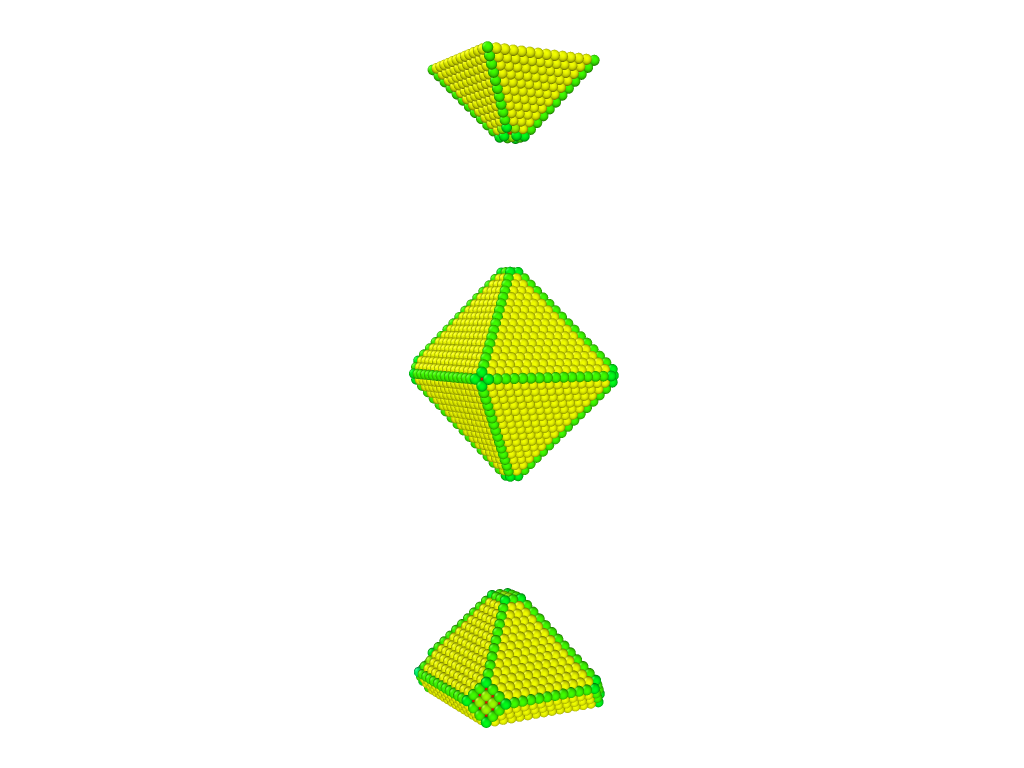}
  }
  \caption{(Colour online) Necking of a 18\;nm long cylindrical $\langle 100 \rangle$ wire at 800\;K due to Rayleigh instability. Periodic boundary conditions are applied, giving the effect of a infinite wire. The wire is shown at 0\;µs (a), 3\;µs (b), and 6\;µs (c). The atoms are coloured according to their coordination number.}
  \label{fig:wires}
  \end{figure*}

\begin{figure}
 \centering
   \includegraphics[width=0.5\textwidth]{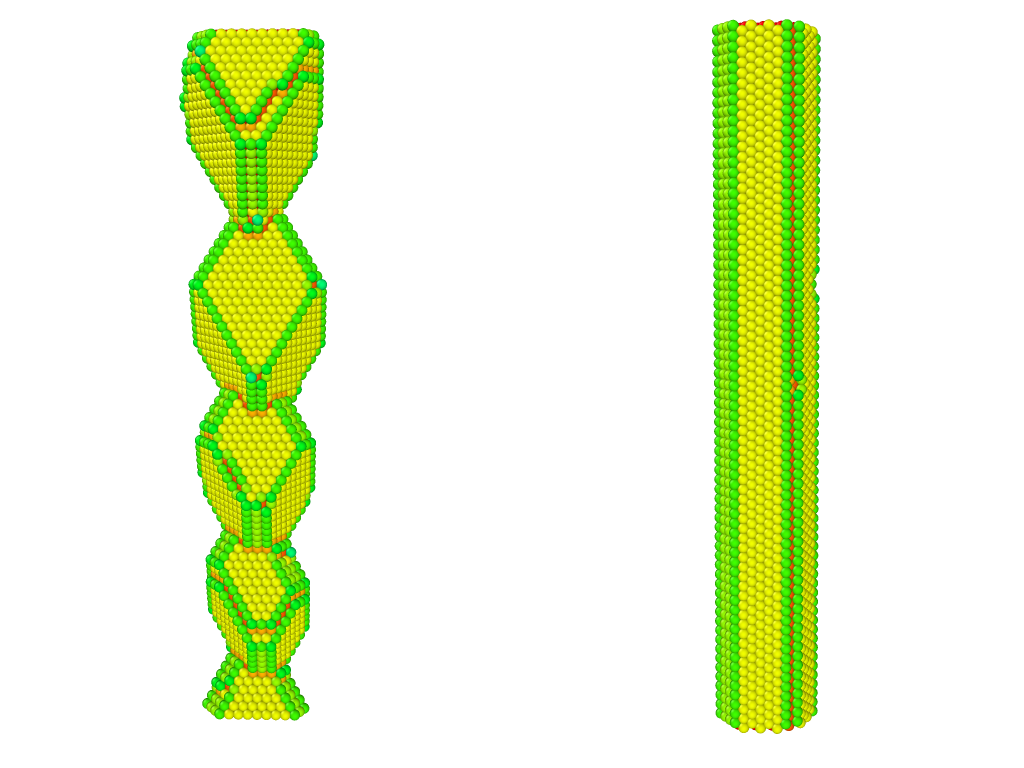}
  \caption{(Colour online) KMC simulations of cylindrical wires in $\langle 111 \rangle$ (left) and $\langle 110 \rangle$ (right) orientation at 800\;K after 10 µs. No necking is observed in either case. Periodic boundary conditions are used. The atoms are coloured according to their coordination number.}
  \label{fig:111_110_wires}
 \end{figure}

We also repeated the simulations with 18\;nm infinitely long cylindrical nanowires; all with a radius $r = 1.1$\;nm. Periodic boundary conditions were applied, giving the effect of infinitely long wires. The wires were directed in the $\langle 100 \rangle$, $\langle 110 \rangle$, and $\langle 111 \rangle$ crystallographic directions. The same necking behaviour was observed for the $\langle 100 \rangle$ wires (figure \ref{fig:wires}) as for the $\langle 100 \rangle$ nanotip, whereas no necking was observed for the $\langle 110 \rangle$ and $\langle 111 \rangle$ wires (figure \ref{fig:111_110_wires}.
 
\begin{figure}
    \centering
    \includegraphics[width=0.5\textwidth]{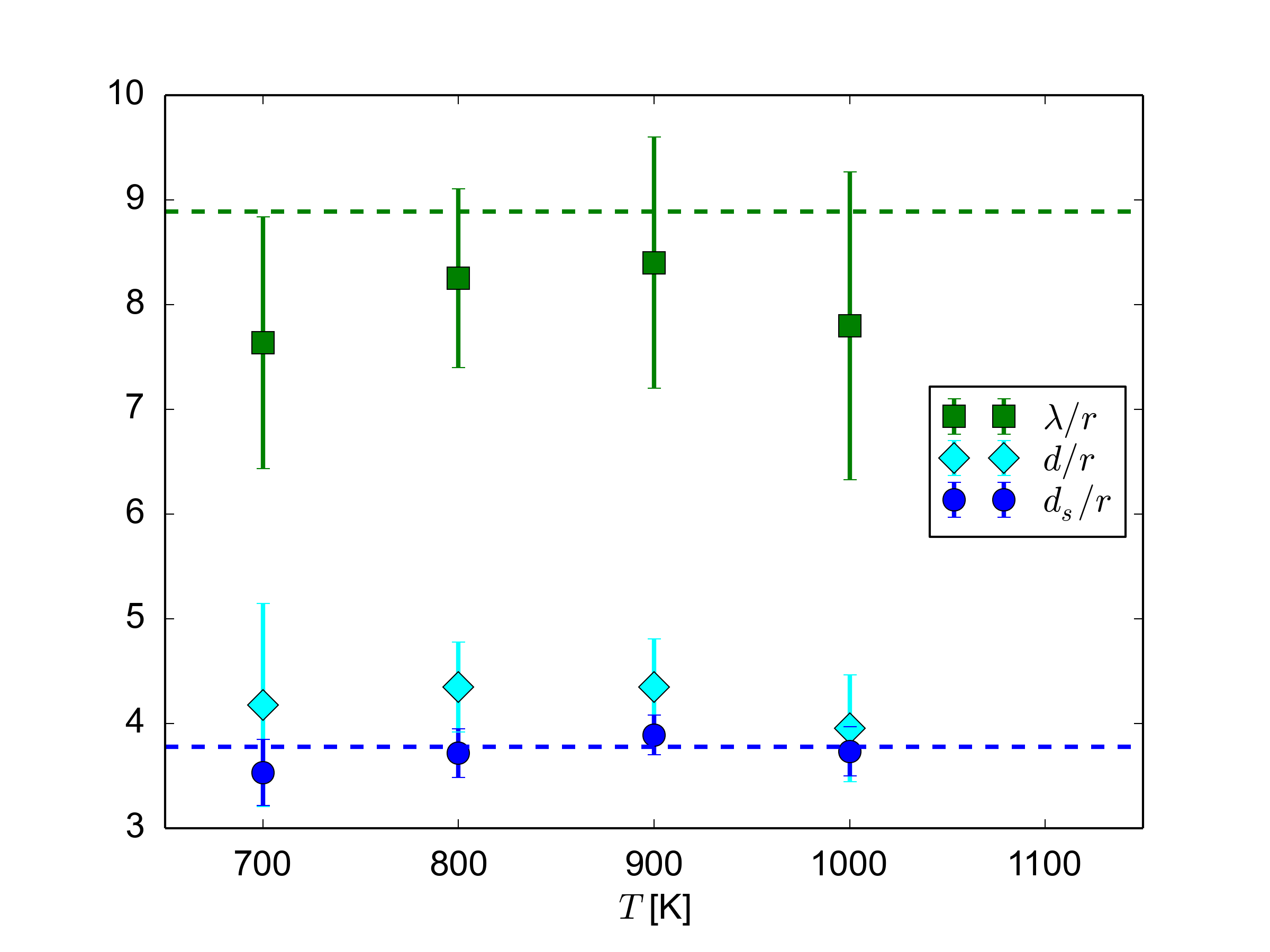}
    \caption{(Colour online) KMC simulations of the breaking up of 18\;nm long cylindrical $\langle 100 \rangle$ wires (with periodic boundary conditions applied) with radius $r$ into clusters with the average diameter $d$ and average spacing $\lambda$. The data is compared with the theoretical values $\lambda/r = 8.89$ and $d/r = 3.78$ \cite{toimil2004fragmentation,nichols1965morphological}. In the simulations, the clusters assume the energetically favourable hexagonal shapes with a diameter $d$, estimated using OVITO \cite{ovito}; if the clusters are approximated as spheres (the number of atoms is conserved) the diameter $d_s$ of such spheres agrees excellently with the theoretical predictions \cite{toimil2004fragmentation,nichols1965morphological}.}
    \label{fig:20151207.png}
\end{figure}

\begin{figure}
    \centering
    \includegraphics[width=0.5\textwidth]{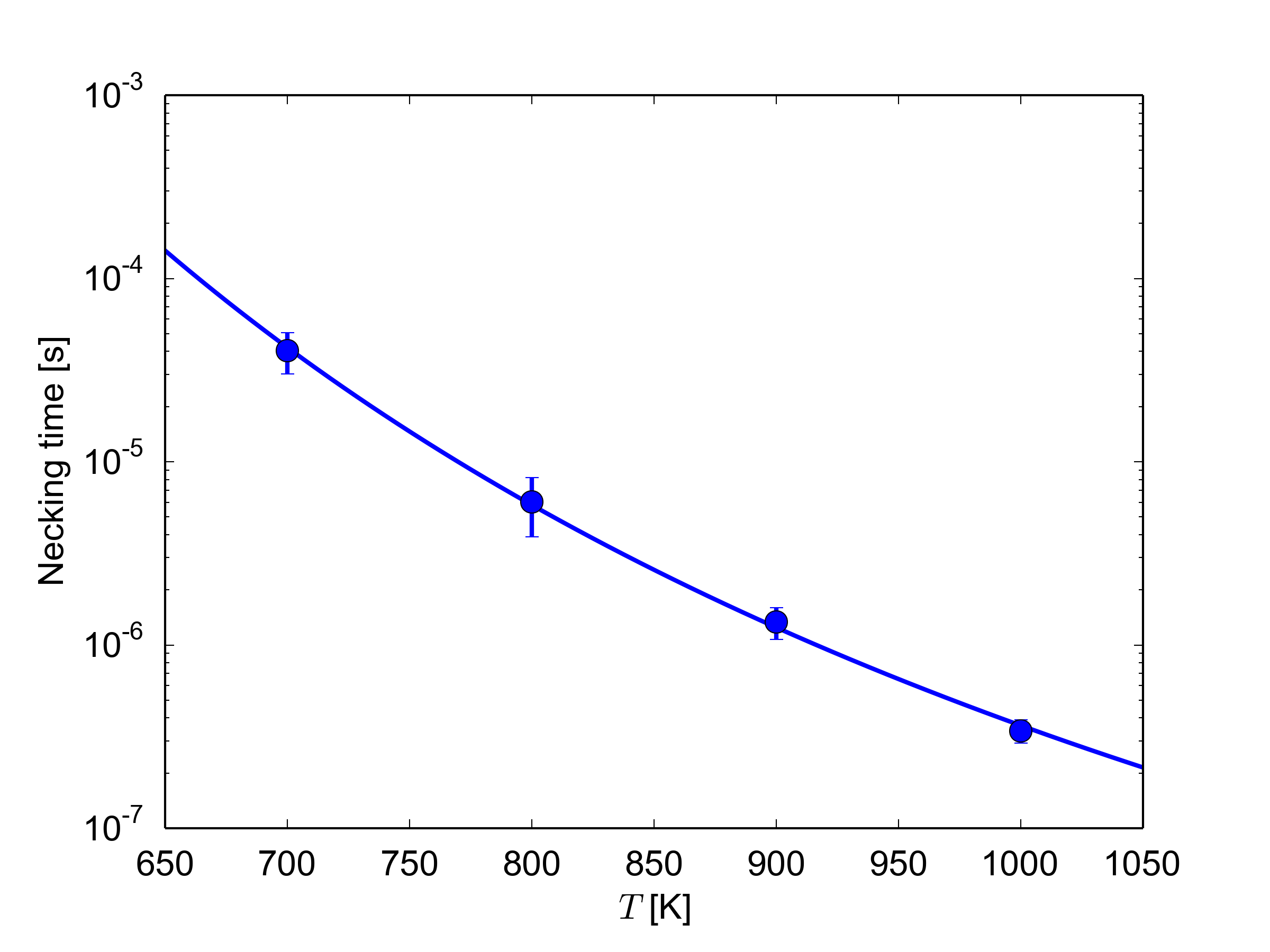}
    \caption{(Colour online) The time until the nanowire disintegrates due to Rayleigh instability as a function of temperature.}
    \label{fig:20151207-t.png}
\end{figure}

For a cylindrical wire with an initial radius $r$, having sinusoidal perturbations resulting in Rayleigh instability, the ratios $\lambda/r = 8.89$ and $d/r = 3.78$ are expected if only surface diffusion is considered \cite{toimil2004fragmentation,nichols1965morphological}. Here, $d$ is the average diameter of the final clusters and $\lambda$ is the average distance between the clusters. The simulations results for the pearling of the wires at different temperatures ranging from 700 to 1000\;K are shown in figure \ref{fig:20151207.png}. If the diameter of the clusters are estimated as spheres with the same volume, good agreement with theory is obtained for both the $d/r$ and $\lambda/r$ ratios. 
The higher ratios of $\lambda/r$ and $d/r$ in figure \ref{fig:20151207.png} are obtained if the diameter of the actual clusters obtained in the simulations are taken into account. The distances $\lambda$ were taken between the centres of the clusters.
No temperature dependence of $d$ and $\lambda$ is observed, in agreement with the simulation results of T\;M\"uller \textit{et al}.\;\cite{muller2002template}. However, the time for the wire to break into clusters do depend on the temperature, as shown in figure \ref{fig:20151207-t.png}. The trend is the same as for the flattening process and can be described by \ref{eq:flattening} using the prefactor $t_0 = 5.60\cdot10^{-12}$\;s and the activation energy $E_a = 0.95$\;eV.

\section{Discussion}\label{sec:discussion}

\subsection{The model}

The key assumptions of our model are that the surface diffusion processes can be correctly described by atom jumps to 1nn positions and that these jumping processes and the associated migration energy, $E_m$, can be characterized solely by the number of 1nn and 2nn atoms of the initial and final positions $(a,b,c,d)$, as described in section \ref{sec:methods}. Jumps to 2nn sites and concerted movements, where more than one atom move at once, are not considered in the model.

The model also assumes all atoms always belong to either the surface or the bulk; jumps to the vacuum above the surface are currently forbidden and evaporation is not considered. Even if clusters detach during the simulation from the main simulation cell, each cluster is considered by \textit{Kimocs} as an independent system; there are no gravitational or other forces implemented that would move a detached cluster as a whole, as would be physically expected. Atoms in the detached cluster may only jump one at a time, as specified by the KMC algorithm.

\subsection{The parameterization of Cu}

The migration barriers for all jump processes were calculated using NEB, which is commonly used for finding the energetically most favourable migration path. Since our parameterization depends on the number of neighbours of the jumping atom, the NEB calculations had to be discarded if any neighbour atoms moved during the relaxation of the system as the resulting barrier would not be calculated for the desired process any more. In particular, we found that atoms with less than 4 atoms in the 1nn positions are very likely to move during the relaxation (and are thus in unstable positions). A similar result was found for aluminium in \cite{huang1998atomistic}, were atoms with less than 3 atoms in the 1nn positions were found to be unstable in atomistic simulations.

All atoms with less than 4 neighbours in the 1nn positions will have zero or near-zero energy barriers for jumps. Although for simplicity, a single small barrier value for these jumps could have been assumed, it may lead to an undesired choice for a jump of the atom which has a very few bonds but still slightly stronger bonded than its neighbouring atom with even less bonds. For instance, having many atoms with few neighbours next to one another might result in an atom jumping to a more bonded position, leaving behind an atom with no nearest neighbours. This will lead to disintegration of the structure and, moreover, violate the principle of KMC. Thus the less stable atom must be given priority to perform the jump.
To cope with this problem, we propose \ref{eq:unstable_atoms}. It is designed to give near-zero barriers for unstable atoms, but the less bonded atoms will, however, have even smaller barriers. This way the integrity of the surface is ensured and the barriers given by \ref{eq:unstable_atoms} do not affect the overall dynamics of the system or the time estimation in \ref{eq:residence_time}.

Since we only consider atom jumps to 1nn positions, it is fairly reasonable to assume the same attempt frequency $\nu$ for all processes. In our model we have fitted the value of $\nu$, which resulted in the same flattening time for the surface nanotip as obtained with MD simulations (section \ref{sec:nu_fitting}). Slightly different flattening times are obtained for different surfaces with MD, as seen in table \ref{table:R20150209}. It has been found in other MD studies that wires with a $\langle 100 \rangle$ orientation will easily undergo a transition to a $\langle 110 \rangle$ orientation  \cite{liang2005shape,park2005shape}. We note that the $\langle 100 \rangle$ to $\langle 110 \rangle$ lattice transition is not possible on a rigid lattice, as in the \textit{Kimocs} model, since a large part of the nanowire
lattice must change orientation in a concerted movement. \textit{Kimocs} will not account for this transition, which explains the large discrepancy between the KMC flattening time for a $\langle 100 \rangle$ oriented nanotip and the MD results. For the fitting of $\nu$, the \{110\} and \{111\} systems are thus the most reliable, and $\nu$ is found to be between $7\cdot10^{13}$\;s$^{-1}$ and $2\cdot10^{14}$\;s$^{-1}$. The chosen value of $\nu = 7\cdot10^{13}$\;s$^{-1}$ is the nearest to the Debye frequency, $4.5\cdot10^{13}$ s$^{-1}$, as often assumed in the KMC community \cite{hook2013solid,soisson2007cu,vincent2008precipitation,castin2011modeling,jansson2013simulation}. This value gives an overestimation of the flattening time with a factor 3 for the $\langle 111 \rangle$ nanotip and a factor 20 for the $\langle 100 \rangle$ nanotip (section \ref{sec:benchmarking}, table \ref{table:R20150209}), which is acceptable.

The KMC data agree well with MD results at temperatures $>$\;800\;K, but the trend differs a bit. However, we do not have MD data to compare with below 850\;K, as the MD method becomes too slow at low temperatures. The extrapolation of the MD data indicates that the flattening time might be much longer than the 3.1\;h predicted by KMC.

By considering the dynamic behaviour of the KMC simulations, we can conclude that the general evolution of the atomic system, as observed in MD simulations, is well reproduced in the case of the flattening of small nanotips (figure \ref{fig:tip_md_kmc}). The coalescence of adatoms into islands, as seen in experiments and other KMC models  \cite{hannon1997surface}, is also correctly reproduced. For the \{111\} surface, one limitation of our model to take into account is that adatoms may not take hcp positions, as discussed in \cite{karim2006diffusion}, as the model only allows fcc positions.

\subsection{The stability of large nanotips}

The KMC simulations of large surface nanotips with aspect ratios of $\sim$7 show that nanotips with the  $\langle 110\rangle$ orientation are particularly stable compared with nanotips of the $\langle 100 \rangle$ or $\langle 111\rangle$ orientations. The latter ones are susceptible to the ``pearling instability''  \cite{toimil2004fragmentation,zhao2006patterning,muller2002template}, that is, breaking into pieces due to Rayleigh necking. In the simulations with the pearling effect, it should be noted that no gravitational nor other external forces, are taken into account;  the pieces detached from the bulk remain suspended in vacuum as an entity in this KMC model.

We have confirmed that the Rayleigh instability is correctly reproduced by our model by simulating the necking of nanowires with good agreement with theory \cite{toimil2004fragmentation,muller2002template,nichols1965morphological} and experiments \cite{toimil2004fragmentation}. Rayleigh instability is observed to occur for the $\langle 100 \rangle$ wires, but not for $\langle 110 \rangle$ and $\langle 111 \rangle$ wires, at temperatures as low as 700\;K. This is much lower than the melting temperature of Cu, even if the finite size effect, that will reduce the melting temperature of Cu nanowires with a thickness of 1.8--2.6\;nm to 900--1000\;K \cite{granberg2014investigation}, would be taken into account. The observed temperature independence of the Rayleigh instability is in good agreement with the results of the KMC studies in \cite{muller2002template}.

Our simulations show that there is no significant difference in the stability of a cuboid nanotip, compared to a cylindrical nanotips with the same height and number of atoms, which can be relevant for the small scale features. It should be noted that the thickness and height of the nanotip will affect the flattening time, as already seen in KMC studies of smaller Cu surface structures by J\;Frantz \textit{et al}.\;\cite{frantz2004evolution}. However, according the our simulations, given a constant room temperature, nanoscale Cu nanotips with aspect ratios even as high as $\sim$7 will be stable for several hours if they have a $\langle110\rangle$ orientation and only diffusion processes are considered.

\section{Conclusions}\label{sec:conclusions}

We have developed a Kinetic Monte Carlo model for the long-term surface evolution of Cu. The model considers atom jumps to first nearest neighbour lattice sites on a rigid lattice. The jumps are characterized by the number of first and second nearest neighbour atoms of the initial and final sites. The KMC model has been found well suited for simulating atomic surface processes and was validated by comparing flattening times of Cu surface nanotips with Molecular Dynamics results for three different surfaces and different temperatures. The computational speed was two orders of magnitude higher with our model than with Molecular Dynamics.

Tips with a $\langle 110\rangle$ orientation were found to be significantly more stable than those with $\langle 111\rangle$ or $\langle 100\rangle$ orientations. Nanowires with a $\langle 110\rangle$ orientation were also found to be stable, as well as wires with the $\langle 111\rangle$ orientation. However, wires with the $\langle 100\rangle$ orientation are found susceptible to the Rayleigh instability, independently of the temperature. The stability of nanotips were found to increase strongly with decreased temperature and a 13\;nm high $\langle 110\rangle$ nanotip with an aspect ratio of $\sim$7 can be expected to be stable for hours at room temperature. However, at temperatures near the melting point, such a nanotip will be reduced to half of its height in less than 100 nanoseconds. The life time of a field emitter in the shape of a nanotip with a large aspect ratio can therefore be assumed to be considerably sensitive to the temperature already by considering the surface diffusion processes alone.

\ack
The authors would like to thank Antti Kuronen for providing his implementation of the NEB method for PARCAS and Kai Nordlund for useful discussions. V\;Jansson was supported by Waldemar von Frenckells Stiftelse, Ruth och Nils-Erik Stenbäcks Stiftelse, and Academy of Finland (Grant No.\;285382). E\;Baibuz was supported by a CERN K-contract. F\;Djurabekova acknowledges gratefully the financial support of Academy of Finland (Grant No.\;269696). Computing resources were provided by the Finnish IT Center for Science (CSC).

\bibliographystyle{iopart-num.bst}
\bibliography{vjansson.bib,vjansson_publications.bib}

\end{document}